\documentclass[10pt,twocolumn,letterpaper]{article}

\usepackage{cvpr}              

\newcommand{\TODO}[1]{\textbf{\color{red}[TODO: #1]}}
\renewcommand{\TODO}[1]{}

\renewcommand{\paragraph}[1]{\vspace{.5em}\noindent\textbf{#1.}}

\usepackage{algorithm}   
\usepackage{algpseudocode} 
\usepackage{xcolor}
\usepackage[accsupp]{axessibility}  

\hfuzz=\maxdimen
\vfuzz=\maxdimen

\definecolor{cvprblue}{rgb}{0.21,0.49,0.74}
\usepackage[pagebackref,breaklinks,colorlinks,allcolors=cvprblue]{hyperref}

\usepackage{graphicx}
\usepackage{subcaption} 
\usepackage{amsmath}  
\usepackage{booktabs} 
\usepackage[table]{xcolor} 
\usepackage[export]{adjustbox}
\usepackage{caption}
\usepackage{array}      
\usepackage{calc}       
\usepackage{rotating}   
\usepackage{multirow}
\definecolor{BestColor}{HTML}{eff3f6}     
\definecolor{BestTextColor}{HTML}{4b6175}    
\definecolor{SecondBestColor}{HTML}{fbfbef}  
\definecolor{SecondBestTextColor}{HTML}{9f7040} 

\def\paperID{32628} 
\def\confName{CVPR}
\def\confYear{2026}

\title{SAGE: Shape-Adapting Gated Experts for Adaptive Histopathology Image Segmentation}

\author{
Gia Huy Thai$^{\ast}$\\
University of Science, VNU-HCM\\
{\tt\small 23120008@student.hcmus.edu.vn}
\and
Hoang-Nguyen Vu$^{\ast}$\\
Trivita AI\\
{\tt\small nguyen.vu@trivita.ai}
\and
Anh-Minh Phan\\
University of Technology, VNU-HCM\\
{\tt\small phananhm5@gmail.com}
\and
Quang-Thinh Ly\\
Michigan State University, USA\\
{\tt\small lythinh@msu.edu}
\and
Tram Dinh\\
Phu Nhuan High School\\
{\tt\small 50.dnmtram.81@gmail.com}
\and
Thi-Ngoc-Truc Nguyen\\
Trivita AI\\
{\tt\small truc.nguyen@trivita.ai}
\and
Nhat Ho\\
The University of Texas at Austin\\
{\tt\small minhnnhat@utexas.edu}
}

\begin{document}
\maketitle

\begingroup
\renewcommand\thefootnote{}\footnotetext{
    {\scriptsize
    $^{\ast}$Equal Contribution.
    }
}
\addtocounter{footnote}{0}
\endgroup

\begin{figure}[t]
    \centering
\scriptsize
\setlength{\belowcaptionskip}{0pt}

\newlength{\imagewidth}
\setlength{\imagewidth}{0.3\columnwidth} 

\newlength{\hsp}
\setlength{\hsp}{0.005em}

\newlength{\vsp}
\setlength{\vsp}{0.4em}

\newlength{\labelvsp}
\setlength{\labelvsp}{0.2em}

\begin{minipage}[t]{\imagewidth}
    \centering
    Original Image \\[\labelvsp]
    \includegraphics[width=0.6\linewidth]{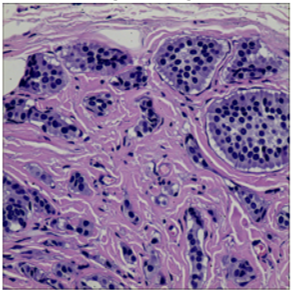} %
\end{minipage}%
\hspace*{\hsp}%
\begin{minipage}[t]{\imagewidth}
    \centering
    Ground Truth \\[\labelvsp]
    \includegraphics[width=0.6\linewidth]{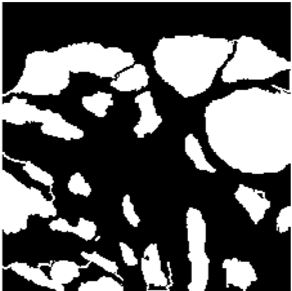}
\end{minipage}%
\hspace*{\hsp}%
\begin{minipage}[t]{\imagewidth}
    \centering
    Prediction \\[\labelvsp]
    \includegraphics[width=0.6\linewidth]{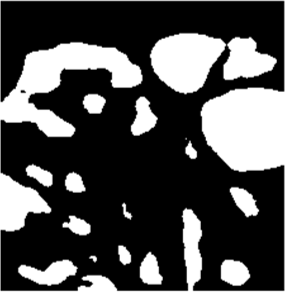}
\end{minipage}%
\\[\vsp]

\begin{minipage}[t]{\imagewidth}
    \centering
    CNN Main Path \\[\labelvsp]
    \includegraphics[width=0.6\linewidth]{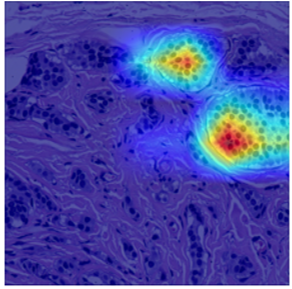}
\end{minipage}%
\hspace*{\hsp}%
\begin{minipage}[t]{\imagewidth}
    \centering
    Transformer Block 1 \\[\labelvsp]
    \includegraphics[width=0.6\linewidth]{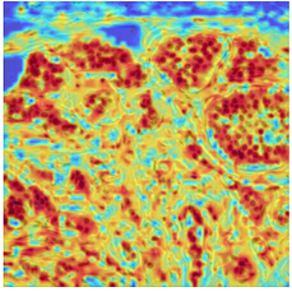}
\end{minipage}%
\hspace*{\hsp}%
\begin{minipage}[t]{\imagewidth}
    \centering
    Transformer Block 11 \\[\labelvsp]
    \includegraphics[width=0.6\linewidth]{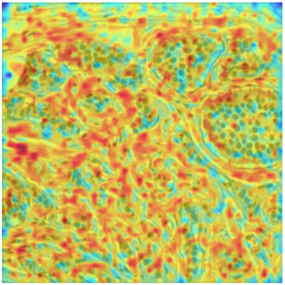}
\end{minipage}%
\\[\vsp]

\begin{minipage}[t]{\imagewidth}
    \centering
    Transformer Main Path \\[\labelvsp]
    \includegraphics[width=0.6\linewidth]{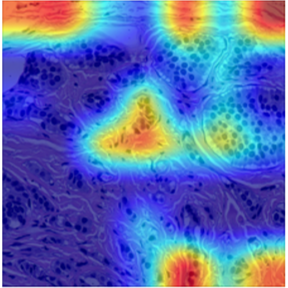}
\end{minipage}%
\hspace*{\hsp}%
\begin{minipage}[t]{\imagewidth}
    \centering
    CNN Block 2 \\[\labelvsp]
    \includegraphics[width=0.6\linewidth]{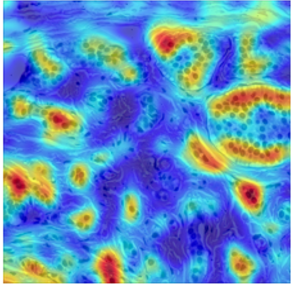}
\end{minipage}%
\hspace*{\hsp}%
\begin{minipage}[t]{\imagewidth}
    \centering
    Transformer Block 2 \\[\labelvsp]
    \includegraphics[width=0.6\linewidth]{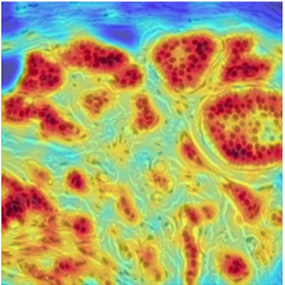}
\end{minipage}


    \caption{\textbf{Explainability visualization of dynamic expert routing in SAGE on the EBHI dataset.} From left to right, the first row shows the input patch, its ground-truth mask, and the predicted segmentation. Rows two and three report representative Grad-CAM heatmaps from the ConvNeXt and Transformer streams, including the main paths and selected expert blocks. Warmer regions indicate higher attribution, illustrating how SAGE routes computation to different experts to refine local boundaries while preserving global context under heterogeneous tissue appearances.}
    \label{fig:visualization_grid}
\end{figure}

\begin{abstract}
The significant variability in cell size and shape continues to pose a major obstacle in computer-assisted cancer detection on gigapixel Whole Slide Images (WSIs), due to cellular heterogeneity. Current CNN-Transformer hybrids use static computation graphs with fixed routing. This leads to extra computation and makes it harder to adapt to changes in input. We propose Shape-Adapting Gated Experts (SAGE), an input-adaptive framework that enables dynamic expert routing in heterogeneous visual networks. SAGE reconfigures static backbones into dynamically routed expert architectures via a dual-path design with hierarchical gating and a Shape-Adapting Hub (SA-Hub) that harmonizes feature representations across convolutional and transformer modules. Embodied as SAGE with ConvNeXt and Vision Transformer UNet (SAGE-ConvNeXt+ViT-UNet), our model achieves a Dice score of $95.23\%$ on EBHI, DSC scores of $92.78\%$ and $91.42\%$ on GlaS Test A and Test B, respectively, and $91.26\%$ DSC at the WSI level on DigestPath, while exhibiting robust generalization under distribution shifts by adaptively balancing local refinement and global context. SAGE establishes a scalable foundation for dynamic expert routing in visual networks, thereby facilitating flexible visual reasoning. 
Project page: \url{https://oxyzgiahuy.github.io/sage/}.
\end{abstract}    
\section{Introduction}
\label{sec:intro}

Computer-aided detection of malignant tissue in gigapixel WSIs is the basis of digital pathology. This makes it possible to quickly and accurately diagnose diseases. For quick diagnosis, classification, and treatment planning of colorectal cancer, it is very important to accurately describe the tumor's morphology. Nonetheless, converting these visually intricate and diverse tissue architectures into computational comprehension continues to be exceedingly difficult. Convolutional Neural Networks (CNNs) \cite{CNN} are great at finding small local features like cell boundaries and textures. Vision Transformers (ViTs) \cite{khan2022transformers}, on the other hand, are a powerful way to model long-range spatial dependencies and global context. Nevertheless, substantial variability in tissue appearance, ranging from homogeneous normal tissues to complex and subtly textured malignant patterns, combined with the large resolution of WSIs, pushes current models beyond their representational and computational limits. Existing models, including U-Net variants and hybrid CNN-Transformer architectures, utilize a static computational graph. This makes all input segments go through the same processing, which is not a good way to do things because it over-processes simple areas and under-models complex ones. Also, the fact that CNN and Transformer blocks can only interact in one way means that you can't take advantage of each paradigm's strengths based on the characteristics of the input.

To address these limitations, we propose \textit{Shape-Adapting Gated Experts} (SAGE), a dynamic, input-adaptive framework that converts a static backbone into a dual-path architecture. Each layer contains a main path that preserves the backbone transformation and an expert path that conditionally activates a subset of reused backbone blocks. A hierarchical router first estimates a group-level preference between shared and fine-grained expert groups, then applies top-$K$ selection on prior-modulated logits to determine the active experts for each input. The two paths are fused adaptively, allowing the model to balance stability and input-specific refinement at run time. To enable interaction across heterogeneous experts (e.g., CNN and Transformer blocks), we introduce the \textit{Shape-Adapting Hub} (SA-Hub), which aligns feature formats before and after expert execution. Although trained on patches, SAGE is deployed on full WSIs through sliding-window reconstruction, preserving compatibility with high-resolution pathology workflows.

To summarize, this work makes contributions as follows:
\begin{itemize}
\item We propose a dual-path formulation that transforms static backbones into dynamically routed architectures, enabling input-adaptive computation with parameter reuse.
\item We design a hierarchical router with group-level gating and top-$K$ selection over prior-modulated logits to balance shared and fine-grained specialization.
\item We introduce SA-Hub, a lightweight shape-adaptation module that aligns CNN/Transformer feature formats for stable cross-expert communication.
\end{itemize}
 
\section{Related Work}
\label{sec:related_work}

\paragraph{Medical Image Segmentation}
Medical image segmentation is a core component of computational pathology because it enables quantitative analysis of cellular and tissue morphology. Earlier methods based on intensity thresholds, region growing, and contour evolution are sensitive to noise and staining variation. Deep learning substantially improved robustness, starting from U-Net~\cite{ronneberger2015unet} and extending to stronger encoders such as ResNet~\cite{ResNet}, EfficientNet~\cite{EfficientNetRM}, ConvNeXt~\cite{ConvNext}, and nested designs such as U-Net++~\cite{zhou2018unet++}. Recent studies emphasize data efficiency and transferability: foundation models such as MedSAM~\cite{medSAM} and SAM-Med2D/3D~\cite{sun2024sammed2d} leverage large-scale pretraining, while semi-supervised methods such as C2GMatch~\cite{C2GMatch} improve performance under limited annotations. However, severe domain heterogeneity in histopathology still makes robust cross-domain adaptation challenging.

\paragraph{Hybrid U-Net Architectures}
Hybrid CNN-Transformer segmentation models seek to combine local detail modeling and long-range context. Representative architectures include TransUNet~\cite{chen2024transunet}, Swin-UNet~\cite{cao2022swin}, and SegFormer~\cite{xie2021segformer}. Although these models improve global reasoning, they still follow static fusion and static computation graphs, which can be suboptimal for highly variable tissue patterns. State-space alternatives such as U-Mamba~\cite{ma2024u} and Swin U-Mamba~\cite{liu2024swin} reduce complexity for large images via linear-time sequence modeling, but fine boundary precision can remain challenging in difficult regions. MoE-based segmentation models (e.g., MoE-NuSeg~\cite{wu2025moe}) introduce conditional computation, yet most existing designs do not explicitly combine depth-wise adaptive routing with cross-architecture shape alignment.

\paragraph{Mixture of Experts}
MoE provides scalable conditional computation by activating only a subset of experts per input~\cite{sparseMoE}. Subsequent advances improve routing robustness and efficiency, including large-scale routing designs~\cite{deepseekMoE} and sigmoid-based gating~\cite{sigmoidGatingMoE, sigmoid_attention}. In vision, MoE has been explored for multimodal routing and multi-task adaptation~\cite{mixtureCNN, Ada-MV}, and decoder-centric parameterization strategies~\cite{mixture-of-low-rank}. Most prior approaches perform routing at token or spatial granularity, with limited depth-wise control over which layers are executed. MoLEx~\cite{molex} addresses this gap by treating layers as experts and routing across depth. SAGE builds on this direction and further introduces hierarchical group-aware routing together with shape-adaptive interaction between heterogeneous experts.
\section{Method}
\label{sec:method}
\begin{figure*}[t]
    \centering
    \includegraphics[width=0.8\linewidth]{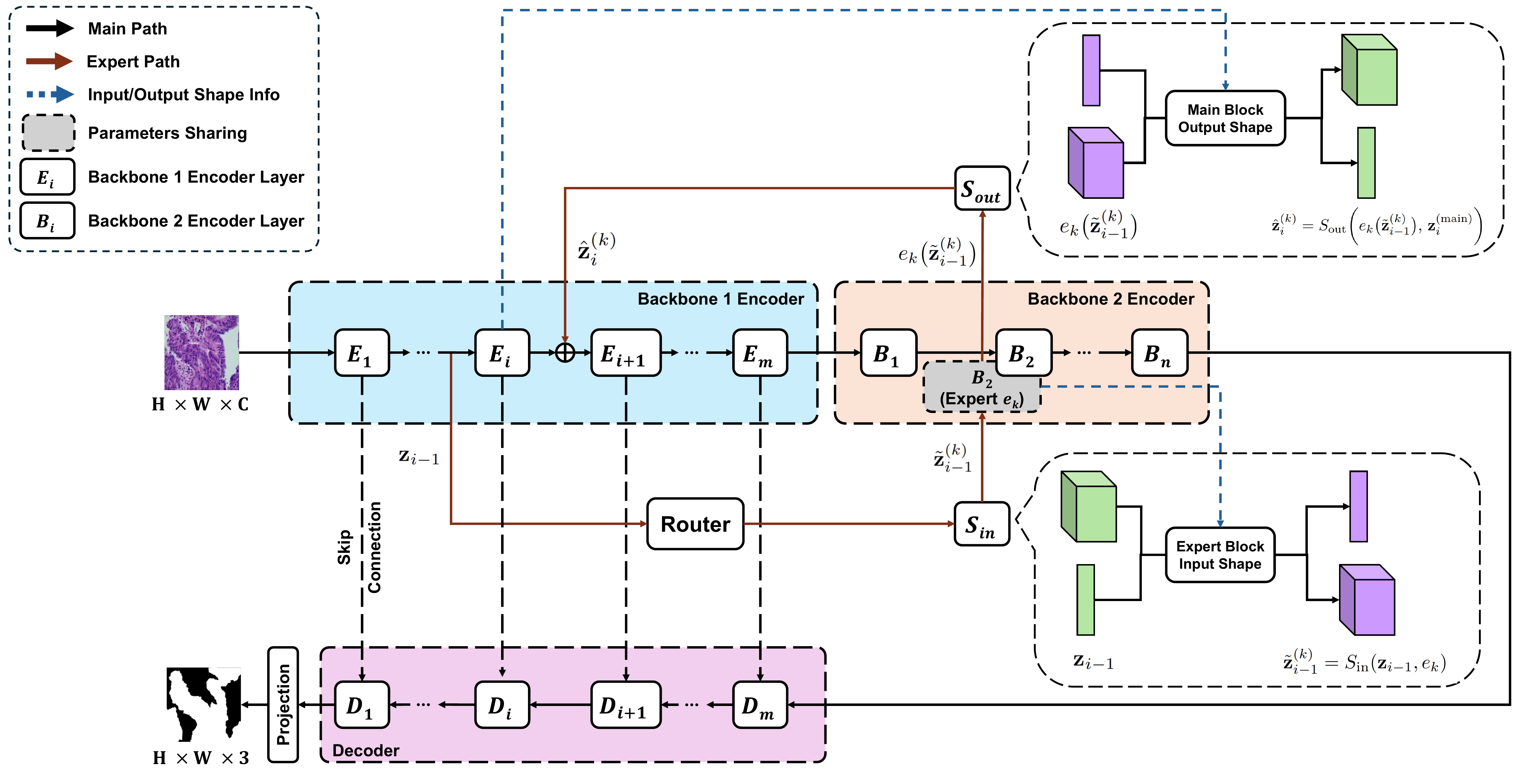}
    \caption{\textbf{Overview of the SAGE framework.} The main path (black arrows) keeps the original forward flow through Backbone 1 and Backbone 2, with multi-scale skip connections to the decoder. In parallel, the expert path (brown arrows) routes features from an intermediate layer to a sparse expert set (illustrated with Stage~2, Expert~$k$) selected by the router. The Shape-Adapting Hub performs bidirectional format alignment via $S_{\text{in}}$ and $S_{\text{out}}$ so that cross-backbone expert execution is shape-compatible before fusion with the main branch. Blue arrows indicate input/output shape constraints used by the adapters, and dashed boxes denote parameter sharing (expert upcycling from pretrained backbone blocks).}
    \label{fig:main-pipeline}
\end{figure*}

Modern hybrid CNN-Transformer architectures generally depend on static computation graphs, applying the same sequence of operations to all inputs regardless of structural intricacy. While this design is stable and easy to implement, it limits adaptability toward heterogeneous visual patterns. To tackle this limitation, Shape-Adapting Gated Experts (SAGE) reparameterizes a fixed backbone into a two-path architecture with conditional expert routing, as illustrated in Figure~\ref{fig:main-pipeline}. This framework preserves the original backbone pathway while introducing sparse expert selection for feature optimization. The Sparse Mixture-of-Experts formulation is introduced in Section~\ref{subsec:preliminaries_moe}; the SAGE block and hierarchical routing are then detailed in Section~\ref{subsec:sage_block_archi} and ~\ref{subsec:hier_routing}. Finally, Section~\ref{subsec:sahub} presents the Shape-Adapting Hub for cross-architecture feature adaptation and integration.

\subsection{Preliminaries: Sparse Mixture-of-Experts}
\label{subsec:preliminaries_moe}
Our SAGE framework is rooted in the Sparse Mixture-of-Experts (SMoE) formulation~\cite{sparseMoE}, which increases capacity by selectively activating only a subset of expert subnetworks, keeping computation roughly constant. A standard SMoE layer comprises a \textit{router} and a set of experts $\{E_j\}_{j=1}^{M}$. Given an input $x$, the SMoE output is
\begin{equation}
y = \sum_{j \in \mathcal{K}} G(x)_j \, E_j(x),
\label{eq:moe_output}
\end{equation}
where $\mathcal{K}$ is the index set of selected experts. There are multiple choices for implementing $G(x)$, but a simple and performant option is to apply a softmax over the Top-$K$ logits of a linear layer, $G(x) := \text{Softmax}(\text{TopK}(x \cdot W_g))$, so only the chosen experts are evaluated. 

During training, a common issue known as \textit{router collapse} arises when only a few experts dominate the routing. 
To promote balanced utilization, an auxiliary \textit{load-balancing loss} encourages uniform token distribution across experts:
\begin{equation}
\mathcal{L}_{\text{load-balancing}} = M \cdot \sum_{j=1}^{M} f_j \, P_j,
\label{eq:balance_loss}
\end{equation}
where $M$ is the total number of experts, $f_j$ denotes the proportion of tokens distributed to expert $j$ and $P_j$ is the proportion of the gating probability assigned to expert $j$. 

Despite its scalability advantages, conventional SMoE primarily decides \emph{which experts} to activate in a single routing stage.
However, many complex tasks require not only expert selection but also adaptive coordination across heterogeneous computation types.
Motivated by this limitation, our SAGE framework generalizes SMoE by introducing hierarchical routing and heterogeneous expert coordination, enabling the model to adaptively determine both \emph{who} computes and \emph{how} computation is performed for each input.

\subsection{SAGE Block Architecture}
\label{subsec:sage_block_archi}
At each backbone layer $i$ (Figure~\ref{fig:main-pipeline}), SAGE replaces a single deterministic transformation with a \emph{dual-path} block that preserves the original computation while enabling learnable refinement. Given an input feature map $\mathbf{z}_{i-1}$, we compute a baseline feature via the original backbone layer and an enriched feature via sparsely activated layer experts, and then fuse them with a learnable gate.

\paragraph{Main path (backbone preservation)}
The main branch applies the original transformation $f_i(\cdot)$ to obtain a stable baseline feature, $\mathbf{z}_i^{(\text{main})} = f_i(\mathbf{z}_{i-1})$.
This pathway anchors optimization and retains the inductive bias and pretrained initialization of the CNN/Transformer backbone.

\paragraph{Expert path (conditional refinement)}
In parallel, the same input is routed to a sparse set of experts. Following MoLEx-style sparse upcycling~\cite{molex}, experts are the pre-trained backbone layers themselves and their parameters are reused rather than replicated; the router activates only a small subset per input. A hierarchical router (Section~\ref{subsec:hier_routing}) performs top-$K$ selection and yields expert weights $\{w_k\}_{k\in\mathcal{K}}$, while the expert-path feature is formed by weighted aggregation into $\mathbf{z}_i^{(\text{expert})}$. The exact construction of $\mathbf{z}_i^{(\text{expert})}$, including shape-adaptive translation and aggregation, is detailed in Section~\ref{subsec:sahub} and Equations~\ref{eq:sahub_in}--\ref{eq:expert_output}. 

\paragraph{Adaptive fusion}
We gate between the baseline and expert-refined features using a learnable scalar $\alpha_i$:
\begin{equation}
    \mathbf{z}_i = \alpha_i \cdot \mathbf{z}_i^{(\text{main})} + (1 - \alpha_i) \cdot \mathbf{z}_i^{(\text{expert})},
    \label{eq:combination}
\end{equation}
where $\alpha_i = \sigma(\theta_i)$ is computed from a learnable parameter $\theta_i$. This formulation allows the model to dynamically balance stability and adaptability, favoring expert-driven refinement when beneficial while preserving the backbone's inductive biases when necessary.

\paragraph{Expert pool composition}
All expert paths draw from a global pool $\mathcal{E}$ with a predefined number of fine-grained experts and shared experts. Both types are implemented as reused backbone layers, but they serve different objectives: fine-grained experts $\mathcal{E}_{\text{fine}}$ focus on depth-specific specialization, whereas shared experts $\mathcal{E}_{\text{shared}}$ encourage domain-generalizable computation. 

\subsection{Hierarchical Expert Routing}
\label{subsec:hier_routing}

\begin{figure}[t]
    \centering    
    \includegraphics[width=0.92 \linewidth]{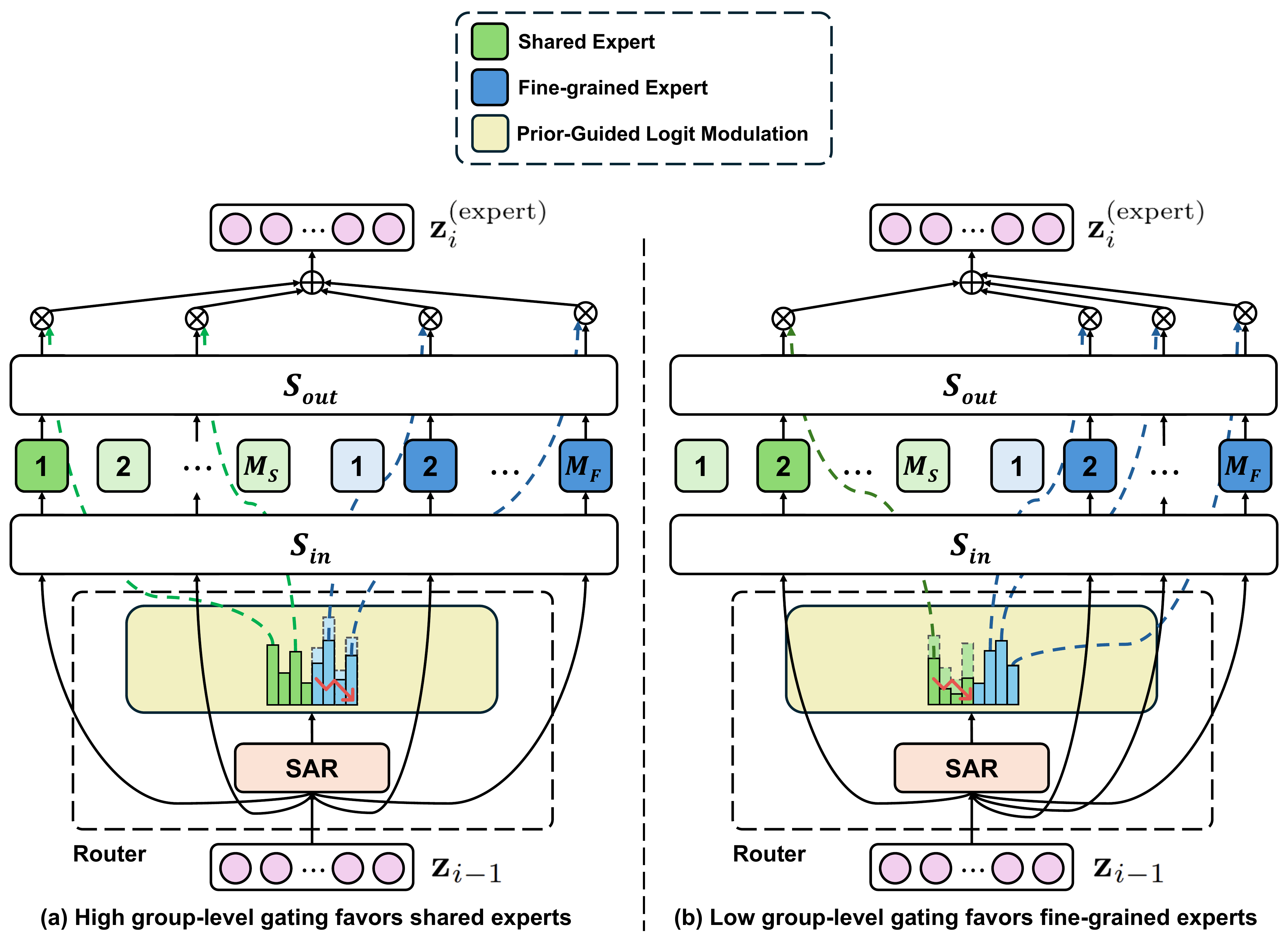}
    \caption{\textbf{Hierarchical routing with prior-guided logit modulation (Top-$K=4$).} \textbf{(a)} For high $g_s$, shared-expert candidates (green) receive a weaker penalty than fine-grained candidates (blue), leading higher Top-$K$ rank. \textbf{(b)} For low $g_s$, the preference reverses, favoring fine-grained experts for input-specific adaptation. Here, $M_S$ and $M_F$ denote the numbers of experts in the shared group and fine-grained group, with total pool size $M=M_S+M_F$; darker fills indicate activated experts. In both regimes, the router computes SAR logits, applies group-prior modulation, performs Top-$K$ selection, and aggregates shape-aligned outputs via $S_{\text{in}}$ and $S_{\text{out}}$ into $\mathbf{z}_i^{(\text{expert})}$.}
    \label{fig:sar}
\end{figure}

SAGE employs a two-level routing strategy to construct a sparse, input-dependent expert path while preserving the backbone computation. As illustrated in Figure~\ref{fig:sar}, given a layer input $\mathbf{z}_{i-1}$, the router computes a group-level gate $g_s$, produces base expert logits via Semantic Affinity Routing (SAR), modulates these logits with the group prior, and then performs top-$K$ selection on the modulated logits. This process jointly controls which experts are preferred (shared versus fine-grained) and which experts are finally executed, yielding the sparse expert computation used in Equation~\ref{eq:combination}.

\paragraph{Group-Level Gating}
A lightweight gating network $G_s$ estimates the group-level preference toward shared experts. It takes a globally pooled representation $\bar{\mathbf{z}}_{i-1} \in \mathbb{R}^d$ and outputs a scalar gate $g_s \in (0,1)$:
\begin{equation}
    g_s = \sigma\left(\bar{\mathbf{z}}_{i-1}\mathbf{W}_{\text{gate}}^{(i)} + b_{\text{gate}}^{(i)}\right).
    \label{eq:gate}
\end{equation}
A high $g_s$ favors shared experts, while a low $g_s$ favors fine-grained experts; shared experts are still selected conditionally via the same top-$K$ routing process.

\paragraph{Semantic Affinity Routing (SAR)}
A primary router $R_i$ computes base logits $\mathbf{L}_i \in \mathbb{R}^{M}$ over all experts $\mathcal{E} = \{E_1, \ldots, E_M\}$:
\begin{equation}
\mathbf{L}_i = \frac{(\bar{\mathbf{z}}_{i-1}\mathbf{W}_{Q}^{(i)})(\mathbf{K}^{(i)})^{\!\top}}{\sqrt{d_k}} + \text{sp}(\bar{\mathbf{z}}_{i-1}\mathbf{W}_{\text{noise}}^{(i)}) \odot \boldsymbol{\epsilon}^{(i)},
\label{eq:sar_logits}
\end{equation}
where $\text{sp}(\cdot)$ is the softplus function, $\boldsymbol{\epsilon}^{(i)} \sim \mathcal{N}(\mathbf{0}, \mathbf{I}_M)$, $\mathbf{W}_{Q}^{(i)} \in \mathbb{R}^{d \times d_k}$ is a learnable query projection, $\mathbf{K}^{(i)} \in \mathbb{R}^{M \times d_k}$ is the expert-key matrix, and $\mathbf{W}_{\text{noise}}^{(i)} \in \mathbb{R}^{d \times M}$ controls input-adaptive noise magnitude. The first term captures semantic affinity, while the second term adds stochastic exploration to improve routing diversity and reduce expert over-specialization. The group-level preference induced by $g_s$ is not applied within SAR; instead, it is introduced afterward via logit modulation.

\paragraph{Prior-Guided Logit Modulation}
Our hierarchical gating couples \emph{group-level} preference (shared vs. fine-grained experts) with \emph{expert-level} selection. Given the base routing logits $\mathbf{L}_i$ produced by SAR and the scalar shared gate $g_s$, we bias expert scores toward the preferred group before selecting the active experts, without forcing shared experts to remain always active. We define a binary mask $\mathbf{m}_s$, where $(\mathbf{m}_s)_j = 1$ if expert $j \in \mathcal{E}_{\text{shared}}$ and $0$ otherwise. The modulated logits are obtained as:
\begin{equation}
    \mathbf{L}'_i = \mathbf{L}_i + \mathbf{m}_s \log(g_s) + (\mathbf{1} - \mathbf{m}_s) \log(1 - g_s),
    \label{eq:logit_modulation}
\end{equation}
where $\mathbf{1} \in \mathbb{R}^M$ is the all-ones vector. This log-space prior raises shared-expert logits when $g_s$ is high and fine-grained logits when $g_s$ is low. For numerical stability, we clip $g_s$ to $[\epsilon, 1-\epsilon]$, then select $\mathcal{K}=\operatorname{TopKIndices}(\mathbf{L}'_i,K)$. Unlike softmax gating, we use independent sigmoid gates on the selected experts:
\begin{equation}
    w_j = \sigma\big((\mathbf{L}'_i)_j\big) \cdot \mathbf{1}[j \in \mathcal{K}],
    \label{eq:topk_sigmoid}
\end{equation}
where $\sigma(\cdot)$ is the sigmoid function; $\mathbf{w} = \{w_j\}_{j=1}^{M}$. For $k \in \mathcal{K}$, $w_k$ denotes the gate value at expert index $k$. This design permits independent multi-expert activation without simplex normalization across selected experts.

Overall, the integration of logit modulation with top-$K$ sigmoid gating induces a sparse, input-adaptive computation graph that balances efficiency with expert specialization across heterogeneous feature distributions.

\begin{algorithm}[!t]
\caption{\textbf{SAGE Training Algorithm (per mini-batch)}}
\label{alg:sage}
\begin{algorithmic}[1]
\Require Input batch $\mathbf{X}$ with labels $\mathbf{Y}$; model $\mathcal{F}$ with $T$ SAGE layers
\Ensure Total loss $\mathcal{L}_{\text{total}}$ for backpropagation
\State $z_0 \gets \text{Stem}(\mathbf{X})$
\State $\mathcal{L}_{\text{load-balancing}} \gets 0$

\For{$i = 1$ to $T$}
    \State $z_i^{(\text{main})} \gets f_i(z_{i-1})$

    \Statex \textbf{Group-Level Gating and SAR}
    \State $\bar{z}_{i-1} \gets \text{GlobalPool}(z_{i-1})$
    \State $g_s \gets \sigma(\bar{z}_{i-1}\mathbf{W}_{\text{gate}}^{(i)} + b_{\text{gate}}^{(i)})$
    \State $\mathbf{L}_i \gets \text{SAR}(\bar{z}_{i-1})$

    \Statex \textbf{Prior-Guided Logit Modulation}
    \State $g_s \gets \text{clip}(g_s, \epsilon, 1-\epsilon)$
    \State $\mathbf{L}'_i \gets \mathbf{L}_i + \mathbf{m}_s\log(g_s) + (\mathbf{1}-\mathbf{m}_s)\log(1-g_s)$

    \Statex \textbf{Top-$K$ Sigmoid Gating and Expert Execution}
    \State $\mathcal{K} \gets \text{TopKIndices}(\mathbf{L}'_i, K)$
    \State $\mathbf{w}[\mathcal{K}] \gets \sigma(\mathbf{L}'_i[\mathcal{K}]);\; \mathbf{w}[\overline{\mathcal{K}}] \gets 0$
    \State $z_i^{(\text{expert})} \gets 0$
    \For{$k \in \mathcal{K}$}
        \State $\tilde{z}_{i-1}^{(k)} \gets S_{\text{in}}(z_{i-1}, e_k)$
        \State $\hat{z}_i^{(k)} \gets S_{\text{out}}(e_k(\tilde{z}_{i-1}^{(k)}), z_i^{(\text{main})})$
        \State $z_i^{(\text{expert})} \gets z_i^{(\text{expert})} + \mathbf{w}_k\,\hat{z}_i^{(k)}$
    \EndFor

    \State $z_i \gets \alpha_i\,z_i^{(\text{main})} + (1-\alpha_i)\,z_i^{(\text{expert})}$
    \State $\mathcal{L}_{\text{load-balancing}} \gets \mathcal{L}_{\text{load-balancing}} + M\sum_{j=1}^{M} f_j^{(i)}P_j^{(i)}$
\EndFor

\State $\mathbf{P} \gets \text{Decoder}(z_T)$
\State $\mathcal{L}_{\text{task}} \gets \lambda_{\text{ce}}\mathcal{L}_{\text{CE}}(\mathbf{P},\mathbf{Y}) + \lambda_{\text{dice}}\mathcal{L}_{\text{Dice}}(\mathbf{P},\mathbf{Y})$
\State $\mathcal{L}_{\text{total}} \gets \mathcal{L}_{\text{task}} + \lambda_{lb}\mathcal{L}_{\text{load-balancing}}$
\State \Return $\mathcal{L}_{\text{total}}$
\end{algorithmic}
\end{algorithm}

\subsection{Cross-Architecture Adaptation and Fusion}
\label{subsec:sahub}
Executing heterogeneous experts requires resolving representation mismatches between convolutional feature maps $(B,C,H,W)$ and token sequences $(B,N,D)$. We address this with the Shape-Adapting Hub, a lightweight learnable module that performs bidirectional format conversion with explicit shape alignment.

\paragraph{Shape-Adapting Hub (SA-Hub)}
SA-Hub consists of an input adapter $S_{\text{in}}$ and an output adapter $S_{\text{out}}$. For each selected expert $e_k$, we first infer the representation type of the source feature and target expert, then apply a format-aware transformation:
\begin{equation}
    \tilde{\mathbf{z}}_{i-1}^{(k)} = S_{\text{in}}(\mathbf{z}_{i-1}, e_k),
    \label{eq:sahub_in}
\end{equation}
with
\begin{equation*}
S_{\text{in}}(\mathbf{x}, e_k)=
\begin{cases}
\mathcal{A}_{C\rightarrow D}\!\left(\mathcal{F}_{\mathrm{tokenize}}(\mathbf{x})\right),
& \begin{array}[t]{@{}l@{}}
\tau(\mathbf{x})=\text{CNN},\\
\tau(e_k)=\text{Trans.},
\end{array}\\[3pt]
\mathcal{F}_{\mathrm{detokenize}}\!\left(\mathcal{A}_{D\rightarrow C}(\mathbf{x})\right),
& \begin{array}[t]{@{}l@{}}
\tau(\mathbf{x})=\text{Trans.},\\
\tau(e_k)=\text{CNN},
\end{array}\\[3pt]
\mathcal{A}(\mathbf{x}), & \tau(\mathbf{x})=\tau(e_k),
\end{cases}
\end{equation*}
where $\tau(\cdot)$ denotes the representation format type (e.g., CNN feature map or Transformer token sequence). $\mathcal{F}_{\mathrm{tokenize}}$ maps spatial feature maps to token sequences, and $\mathcal{F}_{\mathrm{detokenize}}$ reconstructs spatial layouts (with interpolation if needed). $\mathcal{A}_{C\rightarrow D}$ and $\mathcal{A}_{D\rightarrow C}$ denote projections between channel and token-embedding spaces, while $\mathcal{A}(\cdot)$ is used when source and target formats already match. After expert computation, the output adapter maps back to the main-path format with target-shape constraints:
\begin{equation}
    \hat{\mathbf{z}}_i^{(k)} = S_{\text{out}}\bigg(e_k\big(\tilde{\mathbf{z}}_{i-1}^{(k)}\big),\, \mathbf{z}_i^{(\text{main})}\bigg),
    \label{eq:sahub_out}
\end{equation}
where $S_{\text{out}}$ applies the inverse format transformation and enforces the spatial resolution and channel dimensionality of $\mathbf{z}_i^{(\text{main})}$. Finally, the adapted expert outputs are combined by gating-weighted aggregation:
\begin{equation}
    \mathbf{z}_i^{(\text{expert})} = \sum_{k \in \mathcal{K}} w_k \cdot \hat{\mathbf{z}}_i^{(k)}.
    \label{eq:expert_output}
\end{equation}
This design allows CNN and Transformer experts to be mixed within a routing layer without imposing a shared native tensor format. In implementation, SA-Hub maintains a registry of pre-initialized adapters for common dimensional conversions to reduce runtime overhead while preserving end-to-end differentiability. The complete execution flow of SAGE, including hierarchical gating, shape adaptation, and loss computation, is summarized in Algorithm~\ref{alg:sage}.
\section{Experiments}
\label{sec:experiments}

\begin{table*}[t]
\centering
\caption{\textbf{Baseline comparison on EBHI and GlaS.} SAGE-ConvNeXt+ViT-UNet consistently outperforms strong baseline backbones across all reported metrics. All scores are percentages (\%). \colorbox{BestColor}{\textcolor{BestTextColor}{\textbf{Best}}} and \colorbox{SecondBestColor}{\textcolor{SecondBestTextColor}{Second}} indicate the best and second-best performance, respectively.}
\label{tab:baseline_results}

\setlength{\tabcolsep}{2pt}
\setlength{\arrayrulewidth}{0.25pt}
\resizebox{\textwidth}{!}{%
\begin{tabular}{l|ccccc|cccccc|cccccc}
\toprule
\textbf{Model} & \multicolumn{5}{c|}{\centering\textbf{EBHI}} & \multicolumn{6}{c|}{\centering\textbf{GlaS (Test A)}} & \multicolumn{6}{c}{\centering\textbf{GlaS (Test B)}} \\
 & \multicolumn{5}{c|}{\centering\textbf{(Adenocarcinoma)}} & \multicolumn{6}{c|}{\centering\textbf{ }} & \multicolumn{6}{c}{\centering\textbf{ }} \\
\cmidrule{2-6} \cmidrule{7-12} \cmidrule{13-18} 
& Acc $\uparrow$ & IoU $\uparrow$ & DSC $\uparrow$ & HD95 $\downarrow$ & B-F1 $\uparrow$ & Acc $\uparrow$ & IoU $\uparrow$ & DSC $\uparrow$ & HD95 $\downarrow$ & B-F1 $\uparrow$ & O-DSC $\uparrow$ & Acc $\uparrow$ & IoU $\uparrow$ & DSC $\uparrow$ & HD95 $\downarrow$ & B-F1 $\uparrow$ & O-DSC $\uparrow$ \\
\midrule
ResNet101-UNet \cite{ronneberger2015unetconvolutionalnetworksbiomedical, he2015deepresiduallearningimage} 
& 91.93 & 89.11 & 94.24 & 46.15 & 55.58 
& 87.52 & 77.75 & 87.49 & 32.37 & 60.11 & 41.68 
& 85.72 & 78.46 & 87.93 & 26.10 & 56.69 & 40.11 \\

ResNet152-UNet \cite{ronneberger2015unetconvolutionalnetworksbiomedical, he2015deepresiduallearningimage} 
& 91.64 & 88.69 & 94.00 & 45.93 & 54.89 
& 87.60 & 77.76 & 87.60 & 33.08 & 55.79 & 33.81 
& 84.28 & 76.73 & 86.83 & 26.14 & 49.10 & 27.69 \\

EfficientNet-B7-UNet \cite{ronneberger2015unetconvolutionalnetworksbiomedical, tan2020efficientnetrethinkingmodelscaling} 
& 91.82 & 89.00 & 94.18 & 46.37 & 54.45 
& 87.92 & 78.77 & 88.12 & 33.57 & 53.26 & 22.99 
& 84.97 & 77.14 & 87.09 & 30.26 & 51.25 & 24.33 \\

ConvNeXt-UNet \cite{ronneberger2015unetconvolutionalnetworksbiomedical, ConvNext} 
& 91.87 & 89.07 & 94.22 & 46.01 & 54.59 
& \cellcolor{SecondBestColor}\textcolor{SecondBestTextColor}{91.91} & \cellcolor{SecondBestColor}\textcolor{SecondBestTextColor}{85.03} & \cellcolor{SecondBestColor}\textcolor{SecondBestTextColor}{91.91} & \cellcolor{SecondBestColor}\textcolor{SecondBestTextColor}{23.39} & 68.03 & 54.31
& 88.22 & 81.68 & 89.91 & 25.85 & 59.37 & 49.17 \\

U-Net++ (ResNet 101) \cite{zhou2018unet++} 
& 91.94 & 89.13 & 94.25 & \cellcolor{SecondBestColor}\textcolor{SecondBestTextColor}{45.51} & 56.00 
& 89.01 & 80.40 & 89.13 & 30.21 & 61.14 & 29.21 
& 86.23 & 79.30 & 88.46 & 26.34 & 53.08 & 19.38 \\

UMamba \cite{ma2024u} 
& 91.51 & 88.55 & 93.93 & 49.79 & 51.66 
& 83.64 & 72.17 & 83.83 & 34.31 & 53.14 & 36.40 
& 84.46 & 77.22 & 87.15 & 26.23 & 54.89 & 45.00 \\

Swin UMamba \cite{liu2024swin} 
& 92.01 & 89.18 & \cellcolor{SecondBestColor}\textcolor{SecondBestTextColor}{94.28} & 48.35 & 55.23 
& 86.57 & 76.09 & 86.42 & 34.93 & 60.59 & 35.46 
& 84.00 & 75.00 & 85.72 & 26.76 & 58.29 & 38.37 \\

Swin U-Net \cite{cao2022swin} 
& 91.97 & 89.10 & 94.24 & 45.73 & 53.06 
& 90.53 & 82.72 & 90.54 & 26.89 & 60.20 & 45.20 
& 86.82 & 79.52 & 88.59 & 24.82 & 54.70 & 43.09 \\

\midrule

ResNet34-UNet \cite{ronneberger2015unetconvolutionalnetworksbiomedical, he2015deepresiduallearningimage}
& 91.72 & 88.63 & 93.88 & 47.41 & 54.67 
& 88.91 & 81.72 & 89.78 & 26.48 & 63.84 & 55.26 
& 87.46 & 80.68 & 88.92 & 24.62 & 57.38 & 49.84 \\

\textbf{SAGE-ResNet34-UNet (Ours)}
& 92.36 & \cellcolor{SecondBestColor}\textcolor{SecondBestTextColor}{89.24} & \cellcolor{SecondBestColor}\textcolor{SecondBestTextColor}{94.28} & 46.05 & \cellcolor{SecondBestColor}\textcolor{SecondBestTextColor}{56.18} 
& 90.37 & 83.34 & 91.06 & \cellcolor{SecondBestColor}\textcolor{SecondBestTextColor}{23.39} & 69.27 & 61.18 
& 89.28 & \cellcolor{SecondBestColor}\textcolor{SecondBestTextColor}{82.11} & \cellcolor{SecondBestColor}\textcolor{SecondBestTextColor}{90.41} & \cellcolor{SecondBestColor}\textcolor{SecondBestTextColor}{21.27} & 63.45 & 56.73 \\

\midrule

ConvNeXt+ViT-UNet 
& \cellcolor{SecondBestColor}\textcolor{SecondBestTextColor}{92.49} & 83.60 & 90.94 & 46.02 & 55.60 
& 91.88 & 84.91 & 91.80 & 24.25 & \cellcolor{SecondBestColor}\textcolor{SecondBestTextColor}{75.21} & \cellcolor{SecondBestColor}\textcolor{SecondBestTextColor}{65.09} 
& \cellcolor{SecondBestColor}\textcolor{SecondBestTextColor}{89.96} & 81.80 & 89.85 & 22.32 & \cellcolor{SecondBestColor}\textcolor{SecondBestTextColor}{67.57} & \cellcolor{SecondBestColor}\textcolor{SecondBestTextColor}{59.93} \\

\textbf{SAGE-ConvNeXt+ViT-UNet (Ours)} 
& \cellcolor{BestColor}\textcolor{BestTextColor}{94.03} & \cellcolor{BestColor}\textcolor{BestTextColor}{90.90} & \cellcolor{BestColor}\textcolor{BestTextColor}{95.23} & \cellcolor{BestColor}\textcolor{BestTextColor}{45.20} & \cellcolor{BestColor}\textcolor{BestTextColor}{58.10} 
& \cellcolor{BestColor}\textcolor{BestTextColor}{92.96} & \cellcolor{BestColor}\textcolor{BestTextColor}{86.62} & \cellcolor{BestColor}\textcolor{BestTextColor}{92.78} & \cellcolor{BestColor}\textcolor{BestTextColor}{19.85} & \cellcolor{BestColor}\textcolor{BestTextColor}{77.91} & \cellcolor{BestColor}\textcolor{BestTextColor}{73.49} 
& \cellcolor{BestColor}\textcolor{BestTextColor}{91.55} & \cellcolor{BestColor}\textcolor{BestTextColor}{84.56} & \cellcolor{BestColor}\textcolor{BestTextColor}{91.42} & \cellcolor{BestColor}\textcolor{BestTextColor}{17.94} & \cellcolor{BestColor}\textcolor{BestTextColor}{70.23} & \cellcolor{BestColor}\textcolor{BestTextColor}{66.67} \\
\bottomrule
\end{tabular}
}%
\setlength{\tabcolsep}{6pt}
\setlength{\arrayrulewidth}{0.4pt}
\end{table*}

\begin{table*}[t]
\centering
\caption{\textbf{SOTA comparison on EBHI and GlaS.} SAGE-ConvNeXt+ViT-UNet achieves the strongest overall performance across metrics, indicating robust generalization across diverse tissue morphologies. All metrics are reported in percentages (\%). \colorbox{BestColor}{\textcolor{BestTextColor}{\textbf{Best}}} and \colorbox{SecondBestColor}{\textcolor{SecondBestTextColor}{Second}} indicate the best and second-best performance, respectively.}
\label{tab:sota_results}
\setlength{\tabcolsep}{2pt}
\scriptsize
\setlength{\arrayrulewidth}{0.25pt}
\resizebox{\textwidth}{!}{%
\begin{tabular}{l|ccccc|cccccc|cccccc}
\toprule
\textbf{Model} & \multicolumn{5}{c|}{\centering\textbf{EBHI}} & \multicolumn{6}{c|}{\centering\textbf{GlaS (Test A)}} & \multicolumn{6}{c}{\centering\textbf{GlaS (Test B)}} \\
 & \multicolumn{5}{c|}{\centering\textbf{(Adenocarcinoma)}} & \multicolumn{6}{c|}{\centering\textbf{ }} & \multicolumn{6}{c}{\centering\textbf{ }} \\
\cmidrule{2-6} \cmidrule{7-12} \cmidrule{13-18} 
& Acc $\uparrow$ & IoU $\uparrow$ & DSC $\uparrow$ & HD95 $\downarrow$ & B-F1 $\uparrow$ & Acc $\uparrow$ & IoU $\uparrow$ & DSC $\uparrow$ & HD95 $\downarrow$ & B-F1 $\uparrow$ & O-DSC $\uparrow$ & Acc $\uparrow$ & IoU $\uparrow$ & DSC $\uparrow$ & HD95 $\downarrow$ & B-F1 $\uparrow$ & O-DSC $\uparrow$ \\

\midrule
SelfReg-UNet \cite{zhu2024selfreg} 
& 91.53 & 88.58 & 93.95 & 45.66 & 53.83 
& 89.36 & 80.34 & 89.10 & 28.57 & 65.82 & 47.17 
& 86.21 & 77.93 & 87.60 & 26.33 & 59.95 & 40.16 \\

Attention U-Net \cite{oktay2018attention} 
& 92.11 & 89.28 & 94.34 & 46.43 & 57.64 
& 90.53 & 82.43 & 90.37 & 65.60 & 45.95 & 59.03 
& 87.57 & 80.29 & 89.07 & 56.51 & 38.87 & 54.77 \\

ConvUNeXt \cite{han2022convunext} 
& 90.73 & 87.66 & 93.43 & 47.93 & 49.86 
& 87.68 & 77.61 & 87.39 & 28.70 & 62.80 & 45.60 
& 88.05 & 81.16 & 89.60 & \cellcolor{SecondBestColor}\textcolor{SecondBestTextColor}{20.46} & 60.58 & 42.96 \\

UCTransNet \cite{wang2022uctransnet} 
& 91.95 & 89.07 & 94.22 & 46.67 & 57.22 
& 87.07 & 76.47 & 86.66 & 29.23 & 62.97 & 49.38 
& 85.45 & 77.14 & 87.10 & 26.66 & 58.51 & 58.00 \\

TransAttUNet \cite{chen2023transattunet} 
& 91.40 & 88.52 & 93.91 & 50.02 & 51.40 
& 91.18 & 83.69 & 91.12 & 24.51 & 72.13 & 66.65 
& \cellcolor{SecondBestColor}\textcolor{SecondBestTextColor}{89.87} & \cellcolor{SecondBestColor}\textcolor{SecondBestTextColor}{83.78} & \cellcolor{SecondBestColor}\textcolor{SecondBestTextColor}{91.18} & 22.41 & 64.27 & 60.86 \\

SegFormer \cite{xie2021segformer}
& 92.62 & 89.93 & 94.70 & \cellcolor{BestColor}\textcolor{BestTextColor}{42.43} & 55.16 
& 91.09 & 83.31 & 90.89 & \cellcolor{SecondBestColor}\textcolor{SecondBestTextColor}{20.72} & 61.65 & 53.72 
& 87.47 & 80.45 & 89.16 & 23.60 & 55.95 & 51.54 \\

EViT-UNet \cite{li2025evit} 
& \cellcolor{SecondBestColor}\textcolor{SecondBestTextColor}{92.80} & \cellcolor{SecondBestColor}\textcolor{SecondBestTextColor}{90.23} & \cellcolor{SecondBestColor}\textcolor{SecondBestTextColor}{94.86} & 45.30 & \cellcolor{BestColor}\textcolor{BestTextColor}{59.99} 
& \cellcolor{SecondBestColor}\textcolor{SecondBestTextColor}{92.70} & \cellcolor{SecondBestColor}\textcolor{SecondBestTextColor}{86.15} & \cellcolor{SecondBestColor}\textcolor{SecondBestTextColor}{92.56} & 21.82 & \cellcolor{SecondBestColor}\textcolor{SecondBestTextColor}{76.54} & \cellcolor{SecondBestColor}\textcolor{SecondBestTextColor}{73.26} 
& 89.61 & 83.62 & 91.08 & 21.24 & \cellcolor{SecondBestColor}\textcolor{SecondBestTextColor}{65.62} & 63.24 \\

CAC-UNet \cite{zhu2021multi} 
& 91.32 & 88.40 & 93.84 & 51.35 & 53.42 
& 88.01 & 78.34 & 87.85 & 31.70 & 64.29 & 53.81 
& 85.69 & 77.52 & 87.33 & 27.87 & 58.11 & 46.02 \\

TransUNet \cite{chen2021transunet} 
& 91.46 & 88.38 & 93.83 & 45.76 & 53.96 
& 91.26 & 83.72 & 91.14 & 23.46 & 63.62 & 67.33 
& 87.30 & 79.98 & 88.87 & 22.77 & 55.63 & \cellcolor{SecondBestColor}\textcolor{SecondBestTextColor}{63.56} \\

\midrule
\textbf{SAGE-ConvNeXt+ViT-UNet (Ours)} 
& \cellcolor{BestColor}\textcolor{BestTextColor}{94.03} & \cellcolor{BestColor}\textcolor{BestTextColor}{90.90} & \cellcolor{BestColor}\textcolor{BestTextColor}{95.23} & \cellcolor{SecondBestColor}\textcolor{SecondBestTextColor}{45.20} & \cellcolor{SecondBestColor}\textcolor{SecondBestTextColor}{58.10} 
& \cellcolor{BestColor}\textcolor{BestTextColor}{92.96} & \cellcolor{BestColor}\textcolor{BestTextColor}{86.62} & \cellcolor{BestColor}\textcolor{BestTextColor}{92.78} & \cellcolor{BestColor}\textcolor{BestTextColor}{19.85} & \cellcolor{BestColor}\textcolor{BestTextColor}{77.91} & \cellcolor{BestColor}\textcolor{BestTextColor}{73.49} 
& \cellcolor{BestColor}\textcolor{BestTextColor}{91.55} & \cellcolor{BestColor}\textcolor{BestTextColor}{84.56} & \cellcolor{BestColor}\textcolor{BestTextColor}{91.42} & \cellcolor{BestColor}\textcolor{BestTextColor}{17.94} & \cellcolor{BestColor}\textcolor{BestTextColor}{70.23} & \cellcolor{BestColor}\textcolor{BestTextColor}{66.67} \\
\bottomrule
\end{tabular}
}%
\setlength{\tabcolsep}{6pt}
\setlength{\arrayrulewidth}{0.4pt}
\end{table*}

\subsection{Datasets and Evaluation Metrics}
\label{subsec:datasets}
We rigorously evaluated the SAGE framework using three established public benchmarks for colorectal histopathology segmentation: \textit{EBHI},  \textit{GlaS}, and \textit{DigestPath}. 

\paragraph{EBHI Dataset}
The Extended Biopsy Histopathological Image (EBHI) dataset\cite{HU2023102534} contains $5,170$ H\&E-stained biopsy samples, each classified into one of six histological subtypes. We focused on the clinically significant \textit{Adenocarcinoma} subset, and selected $795$ images for our experiments. This dataset evaluates SAGE on heterogeneous yet domain-consistent tissue patterns.

\paragraph{GlaS Dataset} 
The Gland Segmentation (GlaS) dataset~\cite{sirinukunwattana2016glandsegmentationcolonhistology} was introduced in the MICCAI $2015$ Gland Segmentation Challenge. It consists of $165$ H\&E-stained histology images at a resolution of $522 \times 775$, each annotated for glandular structures. The official split includes $85$ images for \textit{Train}, $60$ for \textit{Test A}, and $20$ for \textit{Test B}. This dataset is widely used to assess a model's ability to capture gland morphology and boundary precision.

\paragraph{DigestPath Dataset}
The DigestPath dataset~\cite{DA2022102485} was introduced in the DigestPath $2019$ Challenge and contains $660$ gigapixel whole-slide images (WSIs) from colonoscopy specimens. To facilitate efficient training, we devised a preprocessing pipeline applied to all WSIs. Each WSI was partitioned into overlapping tiles of size $1536 \times 1536$ with stride $512$. A patch $P$ was retained only when both conditions were satisfied on its grayscale intensities:
\begin{equation*}
    P \text{ is retained iff } 
\begin{cases}
\sigma(P) \ge 10, \\
\mu(P) \le 230,
\end{cases}
\end{equation*}
where $\sigma(P)$ and $\mu(P)$ denote the standard deviation and mean intensity of $P$, respectively.
This preprocessing yielded approximately $40{,}000$ patches across all WSIs, filtering out low-information background while preserving diagnostically relevant tissue regions.

\paragraph{Evaluation Metrics}
We assessed segmentation performance using a comprehensive set of metrics: pixel-wise Accuracy (ACC), Intersection over Union (IoU), Dice Similarity Coefficient (DSC), 95\% Hausdorff Distance (HD95), and Boundary F1 (B-F1). For GlaS, given its focus on object-level segmentation rather than semantic segmentation, we additionally reported the Object DSC score (O-DSC).
ACC, IoU, DSC, B-F1, and O-DSC are reported in percentage (\%), while HD95 is reported in pixels.
For EBHI and GlaS, metrics are computed per image and averaged over the held-out test set.
For DigestPath, we report both patch-level and WSI-level results.
To reconstruct WSI-level predictions, overlapping patch logits are mapped back to their original slide coordinates and merged by averaging in overlap regions, followed by a threshold of $0.5$ to obtain the final binary mask; this reconstruction uses the same tiling stride ($512$) as preprocessing.

\subsection{Implementation Details}
\label{subsec:implementation}

\paragraph{SAGE-ConvNeXt+ViT-UNet Configuration}
The SAGE module was integrated into the hybrid encoder combining ConvNeXt \cite{ConvNext} and an ImageNet-pretrained ViT \cite{dosovitskiy2021imageworth16x16words}. The MoE module included $4$ shared experts and $16$ non-shared experts, with the top $4$ experts dynamically selected during routing. We trained SAGE-ConvNeXt+ViT-UNet on two NVIDIA H100 GPUs ($80$GB VRAM each) with a global batch size of $64$ ($32$ per GPU). Training proceeded in two stages using the AdamW optimizer \cite{loshchilov2019decoupledweightdecayregularization} and the hybrid loss $\mathcal{L}_{\text{total}}$ in Algorithm~\ref{alg:sage}, weighted as $\lambda_{\text{ce}} = 1$, $\lambda_{\text{dice}} = 1.5$, $\lambda_{\text{lb}} = 1$. In the first stage, all parameters were optimized with a uniform learning rate of $1\times10^{-5}$. In the second stage, \textit{discriminative fine-tuning} \cite{howard2018universallanguagemodelfinetuning} was applied with $1\times10^{-5}$ for non-shared experts, routers, and the decoder, and $5\times10^{-5}$ for shared experts. Unless noted otherwise, all experiments used random seed $42$, and other hyperparameters followed default settings. 

\paragraph{Comparison Protocol}
To benchmark SAGE-ConvNeXt+ViT-UNet, we compared its performance in two settings: \textit{Baseline Comparison} and \textit{State-of-the-Art (SOTA) Comparison}. For GlaS and EBHI, we report results under both settings, with all input images resized to $224 \times 224$. In contrast, DigestPath was evaluated only under the SOTA setting, using a higher resolution of $512 \times 512$ to preserve critical tissue-level detail. We adhered to the official training setups for all methods (e.g., optimizer, learning rate), except for a standardized batch size of $64$, which ensured efficient multi-GPU utilization.

\paragraph{Evaluation and Model Selection} Dataset splitting strategies varied by dataset structure. For GlaS, we split the official training data into $80\%$ for training and $20\%$ for validation, using the official Test A and Test B sets for final evaluation. For EBHI, we employed a random split of $70\%$ training, $15\%$ validation, and $15\%$ testing at the image level. For DigestPath, we performed a stratified split of $70\%/15\%/15\%$ based on the raw positive and negative WSIs. The patch extraction pipeline was subsequently applied to the WSIs within each split. For all datasets, the model checkpoint with the highest Dice score on the validation set was selected for final evaluation on the test set.

\subsection{Quantitative Comparison}

This section provides a quantitative comparison of SAGE-ConvNeXt+ViT-UNet against strong baselines and SOTA methods. Baseline results are in Table~\ref{tab:baseline_results}, while SOTA comparisons are in Table~\ref{tab:sota_results} and Table~\ref{tab:digestpath_sota_results}.

\begin{table}[htp]
\centering
\caption{\textbf{SOTA comparison on DigestPath at Patch and WSI levels.} SAGE-ConvNeXt+ViT-UNet delivers the best overall results under both evaluation protocols. All metrics are reported in percentages (\%). \colorbox{BestColor}{\textcolor{BestTextColor}{\textbf{Best}}} and \colorbox{SecondBestColor}{\textcolor{SecondBestTextColor}{Second}} indicate the best and second-best performance among all models, respectively.}
\label{tab:digestpath_sota_results}

\setlength{\tabcolsep}{2pt}
\scriptsize
\setlength{\arrayrulewidth}{0.25pt}
\resizebox{\columnwidth}{!}{%
\begin{tabular}{l|ccccc|ccccc}
\toprule
\textbf{Model} & \multicolumn{10}{c}{\centering\textbf{DigestPath}} \\
\cmidrule{2-11}
& \multicolumn{5}{c|}{\centering\textbf{Patch}} & \multicolumn{5}{c}{\centering\textbf{WSI}} \\
\cmidrule{2-6} \cmidrule{7-11}
& Acc $\uparrow$ & IoU $\uparrow$ & DSC $\uparrow$ & HD95 $\downarrow$ & B-F1 $\uparrow$ & Acc $\uparrow$ & IoU $\uparrow$ & DSC $\uparrow$ & HD95 $\downarrow$ & B-F1 $\uparrow$ \\
\midrule
SelfReg-UNet \cite{zhu2024selfreg}
& 96.28 & 80.05 & 83.32 & 131.20 & 71.61
& 97.77 & 78.06 & 83.04 & 417.45 & 60.09 \\

Attention U-Net \cite{oktay2018attention}
& 96.81 & 80.47 & 83.97 & 123.71 & 72.29
& \cellcolor{SecondBestColor}\textcolor{SecondBestTextColor}{98.03} & 80.30 & 85.36 & 416.23 & 61.31 \\

ConvUNeXt \cite{han2022convunext}
& 96.84 & 77.71 & 81.22 & 123.98 & 69.40
& \cellcolor{SecondBestColor}\textcolor{SecondBestTextColor}{98.03} & 74.20 & 79.25 & 576.06 & 55.37 \\

UCTransNet \cite{wang2022uctransnet}
& 96.63 & 81.54 & 84.89 & 121.27 & 73.81
& 97.95 & 84.44 & 89.44 & 427.20 & 66.35 \\

TransAttUNet \cite{chen2023transattunet}
& 96.86 & 80.37 & 83.69 & 126.87 & 71.37
& 98.02 & 75.53 & 80.45 & 454.48 & 56.00 \\

SegFormer \cite{xie2021segformer}
& \cellcolor{SecondBestColor}\textcolor{SecondBestTextColor}{96.99} & 84.54 & 87.66 & 115.75 & 76.54
& 97.97 & \cellcolor{SecondBestColor}\textcolor{SecondBestTextColor}{85.51} & \cellcolor{SecondBestColor}\textcolor{SecondBestTextColor}{90.56} & 393.11 & \cellcolor{SecondBestColor}\textcolor{SecondBestTextColor}{67.15} \\

EViT-UNet \cite{li2025evit}
& 96.60 & 83.38 & 86.37 & 120.47 & 76.38
& 98.02 & 83.16 & 87.78 & \cellcolor{SecondBestColor}\textcolor{SecondBestTextColor}{364.86} & 65.97 \\

CAC-UNet \cite{zhu2021multi}
& 96.81 & 81.74 & 84.99 & \cellcolor{SecondBestColor}\textcolor{SecondBestTextColor}{61.10} & \cellcolor{SecondBestColor}\textcolor{SecondBestTextColor}{78.78}
& 97.96 & 83.63 & 88.66 & 528.06 & 65.03 \\

TransUNet \cite{chen2021transunet}
& 96.79 & 83.19 & 86.50 & 119.96 & 74.40
& 97.92 & 82.25 & 87.28 & 419.77 & 63.18 \\

\midrule

ConvNeXt+ViT-UNet
& 96.80 & \cellcolor{SecondBestColor}\textcolor{SecondBestTextColor}{89.51} & \cellcolor{SecondBestColor}\textcolor{SecondBestTextColor}{91.96} & 132.88 & 75.91
& 97.95 & 83.46 & 88.69 & 482.08 & 65.16 \\

\textbf{SAGE-ConvNeXt+ViT-UNet (Ours)}
& \cellcolor{BestColor}\textcolor{BestTextColor}{97.69} & \cellcolor{BestColor}\textcolor{BestTextColor}{90.21} & \cellcolor{BestColor}\textcolor{BestTextColor}{92.66} & \cellcolor{BestColor}\textcolor{BestTextColor}{60.67} & \cellcolor{BestColor}\textcolor{BestTextColor}{79.48}
& \cellcolor{BestColor}\textcolor{BestTextColor}{98.73} & \cellcolor{BestColor}\textcolor{BestTextColor}{86.21} & \cellcolor{BestColor}\textcolor{BestTextColor}{91.26} & \cellcolor{BestColor}\textcolor{BestTextColor}{362.31} & \cellcolor{BestColor}\textcolor{BestTextColor}{67.85} \\
\bottomrule
\end{tabular}
}%
\setlength{\tabcolsep}{6pt}
\setlength{\arrayrulewidth}{0.4pt}
\end{table}

\paragraph{Baseline Comparison}
Table~\ref{tab:baseline_results} demonstrates that SAGE enhances both pure-CNN and hybrid backbones, rather than benefiting a single architecture. In a direct plug-in comparisons, SAGE-ResNet34-UNet improves over ResNet34-UNet on EBHI and on both GlaS splits. The gains are especially clear on gland-level structure quality, where O-DSC increases by about $+5.9\%$/$+6.9\%$ on Test A/B. A similar trend is observed in the hybrid pair: SAGE-ConvNeXt+ViT-UNet also outperforms its non-SAGE counterpart on EBHI, with strong gains in IoU and DSC (approximately $+7.30\%$ IoU and $+4.29\%$ DSC). Overall, these paired comparisons indicate that the performance gain is mainly attributable to dynamic routing, not to a particular encoder design.

\paragraph{SOTA Comparison} Tables~\ref{tab:sota_results} and~\ref{tab:digestpath_sota_results} present a comparison between the proposed model and recent SOTA methods. SAGE-ConvNeXt+ViT-UNet achieves the highest overall performance across EBHI, GlaS, and DigestPath, with the most significant margins observed in more challenging settings. On EBHI and GlaS, it surpasses strong transformer-based competitors while maintaining the highest DSC values. On DigestPath, it ranks first at both patch and WSI levels, with a notably larger gain at WSI level ($+2.57\%$ DSC) compared to the patch level ($+0.70\%$ DSC). This pattern aligns with the objective of enhancing robustness to morphological variation and domain shift.

\subsection{Efficiency and Model Complexity}
\label{subsec:complexity_main}

Table~\ref{tab:complexity_main} highlights two key points. First, despite adding dual-path routing and SA-Hub adapters, SAGE keeps parameter growth modest ($543.71$M $\rightarrow$ $573.51$M, $+5.5\%$) because expert blocks are reused from the backbone. Second, computation scales with Top-$K$: FLOPs increase from $63.77$G (baseline) to $99.51$G ($K=1$), $130.47$G ($K=2$), and $486.81$G ($K=4$), reflecting the expected accuracy--efficiency trade-off of sparse expert activation.

\subsection{Qualitative Results}
\label{subsubsec:qualitative}

\begin{figure}[t]
    \centering
    \setlength{\tabcolsep}{1.2pt} 
    \renewcommand{\arraystretch}{0.91}

    \newcommand{\imgw}{0.138\textwidth}
    \newcommand{\stackvsep}{0.4mm}

    \resizebox{\linewidth}{!}{%
\begin{tabular}{@{}p{0.015\textwidth}cccc@{}}
        & \scriptsize\textbf{(a) Input} &
        \scriptsize\textbf{(b) TransUNet} &
        \scriptsize\textbf{(c) EViT-UNet} &
        \scriptsize\textbf{(d) Ours} \\[2pt]

        \raisebox{-0.5\totalheight}{\rotatebox{90}{\scriptsize\textbf{GlaS Test A}}} &
        \begin{minipage}{\imgw}
            \includegraphics[width=\linewidth]{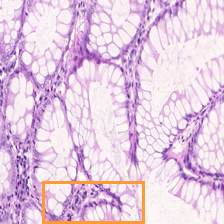}\\[-\stackvsep]
            \includegraphics[width=\linewidth]{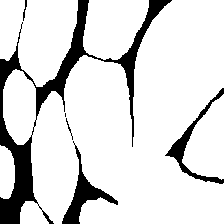}
        \end{minipage} &
        \begin{minipage}{\imgw}
            \includegraphics[width=\linewidth]{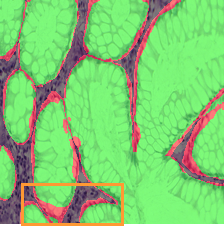}\\[-\stackvsep]
            \includegraphics[width=\linewidth]{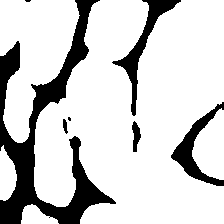}
        \end{minipage} &
        \begin{minipage}{\imgw}
            \includegraphics[width=\linewidth]{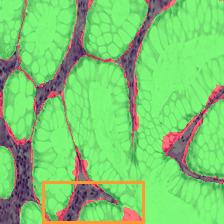}\\[-\stackvsep]
            \includegraphics[width=\linewidth]{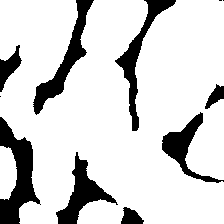}
        \end{minipage} &
        \begin{minipage}{\imgw}
            \includegraphics[width=\linewidth]{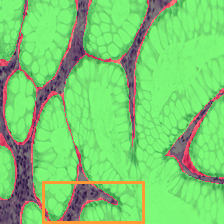}\\[-\stackvsep]
            \includegraphics[width=\linewidth]{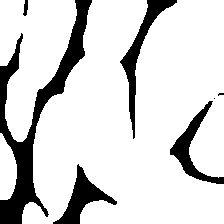}
        \end{minipage} \\[2pt]

        \raisebox{-0.5\totalheight}{\rotatebox{90}{\scriptsize\textbf{GlaS Test B}}} &
        \begin{minipage}{\imgw}
            \includegraphics[width=\linewidth]{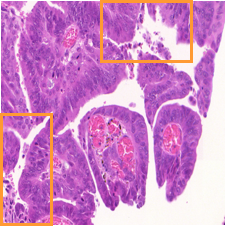}\\[-\stackvsep]
            \includegraphics[width=\linewidth]{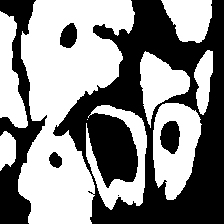}
        \end{minipage} &
        \begin{minipage}{\imgw}
            \includegraphics[width=\linewidth]{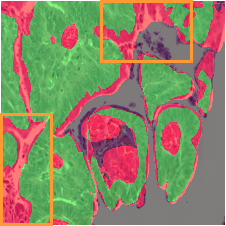}\\[-\stackvsep]
            \includegraphics[width=\linewidth]{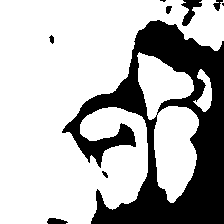}
        \end{minipage} &
        \begin{minipage}{\imgw}
            \includegraphics[width=\linewidth]{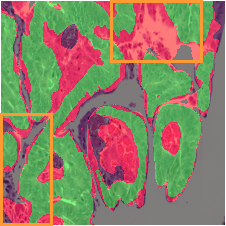}\\[-\stackvsep]
            \includegraphics[width=\linewidth]{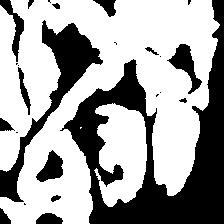}
        \end{minipage} &
        \begin{minipage}{\imgw}
            \includegraphics[width=\linewidth]{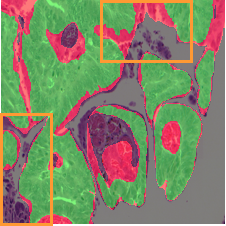}\\[-\stackvsep]
            \includegraphics[width=\linewidth]{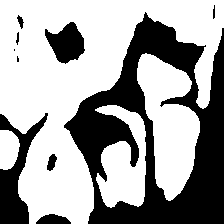}
        \end{minipage}
    \end{tabular}%
}

    \caption{\small
        \textbf{Qualitative comparison on GlaS test samples.}
        Each column shows (a) the input image with ground-truth annotation, (b) TransUNet, (c) EViT-UNet, and (d) our proposed SAGE-ConvNeXt+ViT-UNet. 
        The top row presents a typical gland structure (GlaS Test A), while the bottom row depicts a challenging case with irregular morphology (GlaS Test B). 
        Green areas denote correct predictions, and red areas denote errors.}
    \label{fig:qualitative_results}
\end{figure}

\begin{table}[t]
\centering
\caption{\textbf{Model complexity across varying Top-$K$ settings.} Parameter count and FLOPs per image for the baseline and SAGE.}
\label{tab:complexity_main}
\setlength{\tabcolsep}{6pt}
\small
\resizebox{\columnwidth}{!}{%
\begin{tabular}{@{}lccc@{}}
\toprule
\textbf{Model} & \textbf{Top-$K$} & \textbf{Params (M)} & \textbf{GFLOPs/image} \\
\midrule
ConvNeXt+ViT-UNet (Baseline) & -- & 543.71 & 63.77 \\
SAGE-ConvNeXt+ViT-UNet & 1 & 573.51 & 99.51 \\
SAGE-ConvNeXt+ViT-UNet & 2 & 573.51 & 130.47 \\
SAGE-ConvNeXt+ViT-UNet & 4 & 573.51 & 486.81 \\
\bottomrule
\end{tabular}%
}
\normalsize
\setlength{\tabcolsep}{6pt}
\end{table}

\begin{figure}[t!]
\centering

\newcommand{\gap}{1pt}              
\setlength{\tabcolsep}{1pt}         
\renewcommand{\arraystretch}{1}

\newcommand{\imgw}{0.138\textwidth} 

\resizebox{\linewidth}{!}{%
\begin{tabular}{@{}p{0.015\textwidth}cccc@{}}
    & \scriptsize\textbf{(a) Input} &
    \scriptsize\textbf{(b) SegFormer} &
    \scriptsize\textbf{(c) EViT-UNet} &
    \scriptsize\textbf{(d) Ours} \\[2pt]

    \raisebox{-0.5\totalheight}{\rotatebox{90}{\scriptsize\textbf{WSI Sample 1}}} &
    \begin{minipage}{\imgw}
        \includegraphics[width=\linewidth]{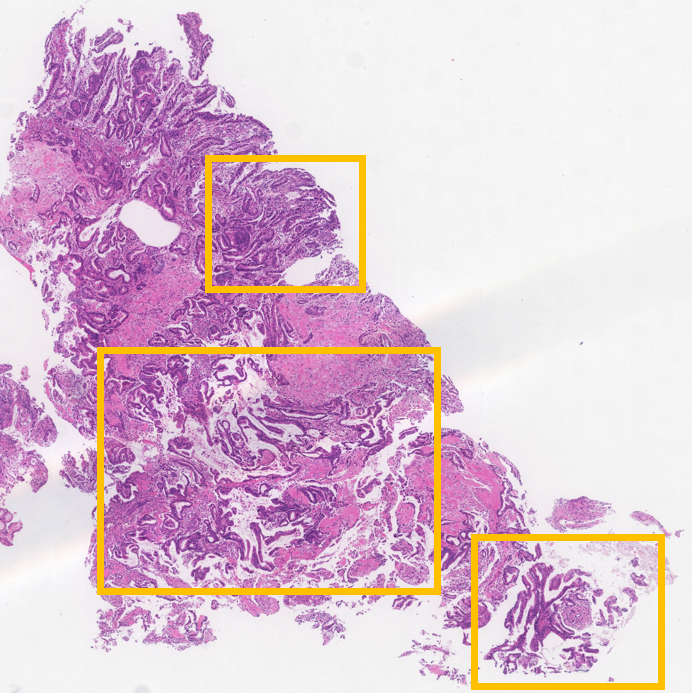}\\[\gap]
        \includegraphics[width=\linewidth]{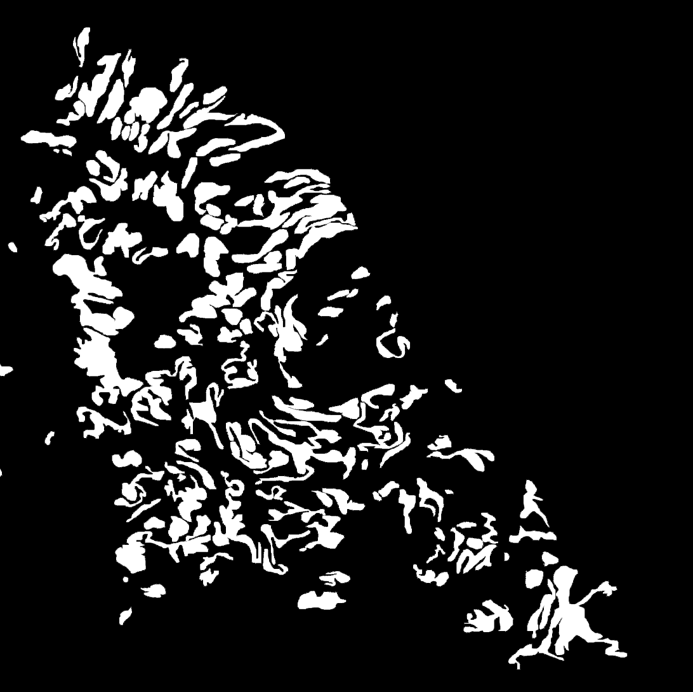}
    \end{minipage} &
    \begin{minipage}{\imgw}
        \includegraphics[width=\linewidth]{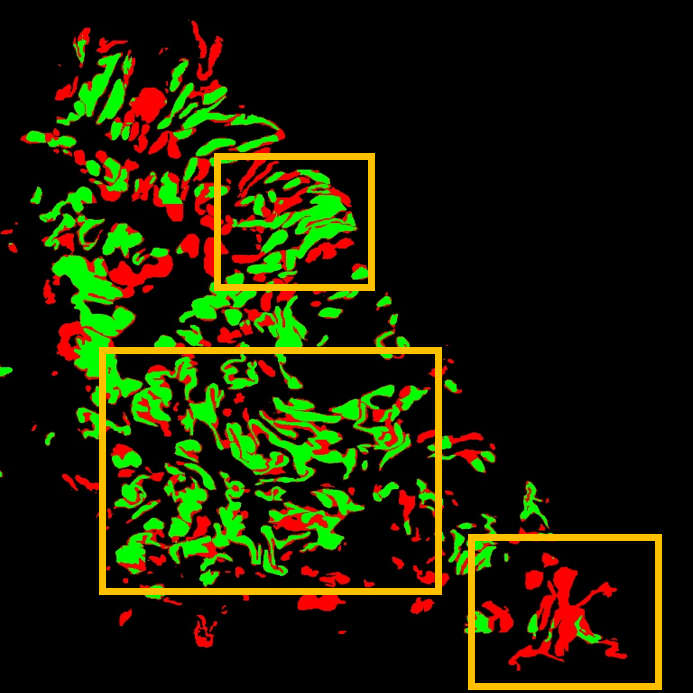}\\[\gap]
        \includegraphics[width=\linewidth]{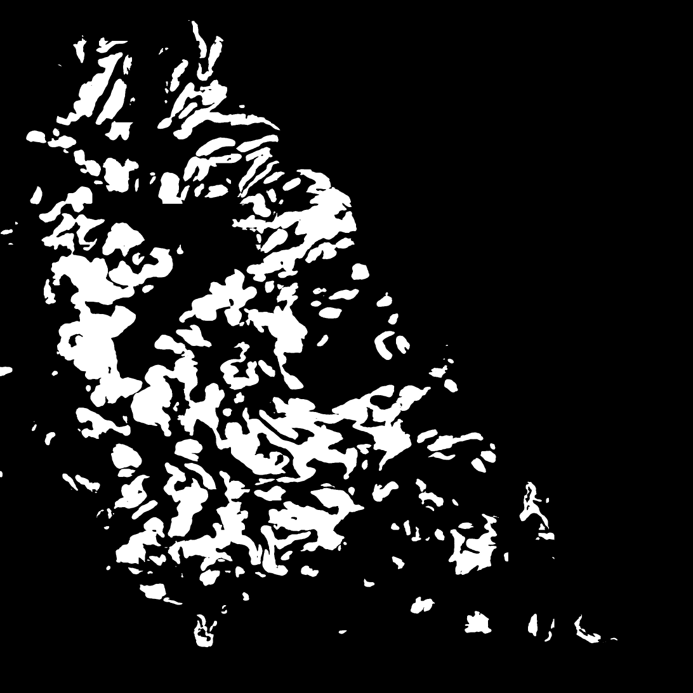}
    \end{minipage} &
    \begin{minipage}{\imgw}
        \includegraphics[width=\linewidth]{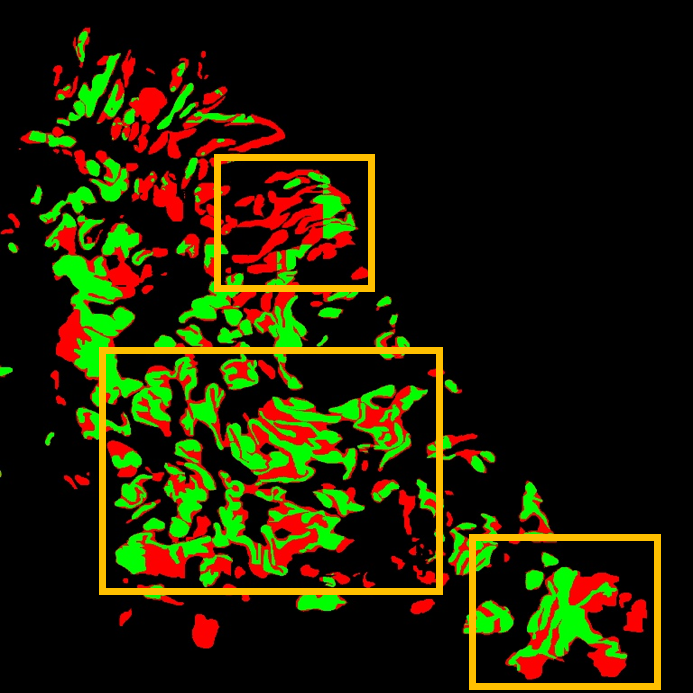}\\[\gap]
        \includegraphics[width=\linewidth]{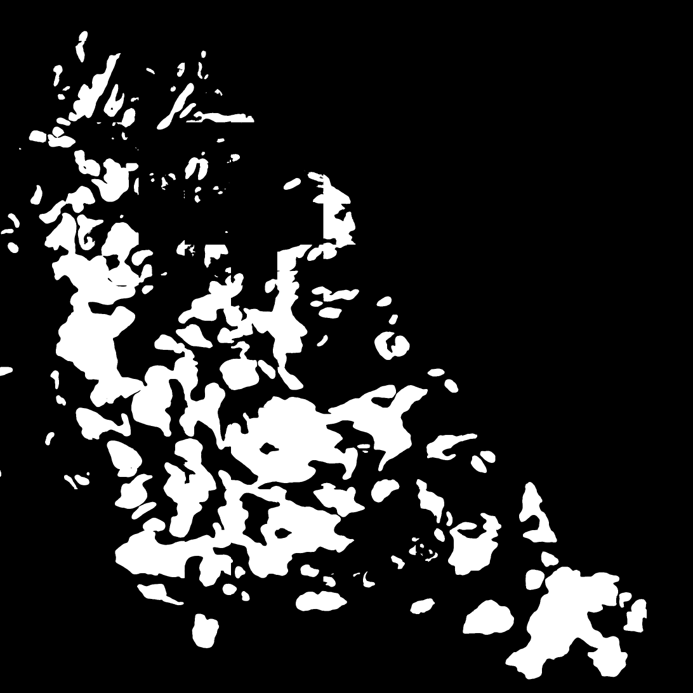}
    \end{minipage} &
    \begin{minipage}{\imgw}
        \includegraphics[width=\linewidth]{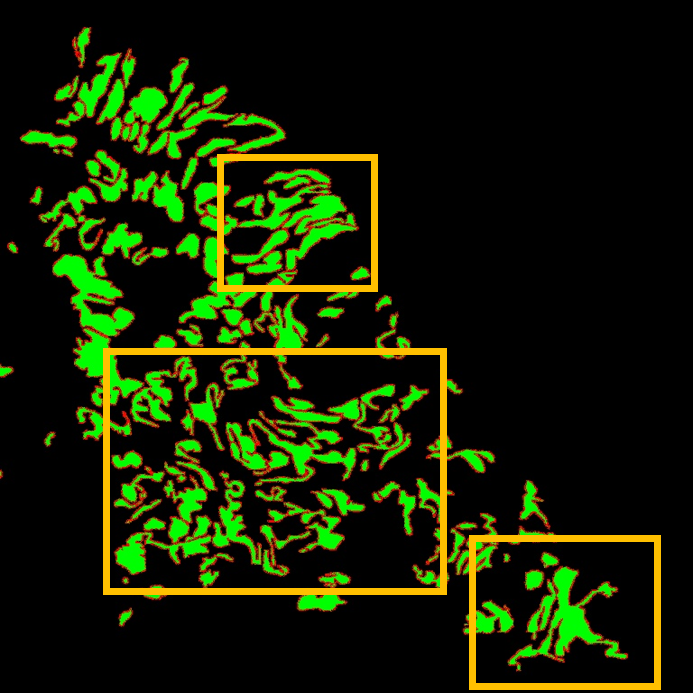}\\[\gap]
        \includegraphics[width=\linewidth]{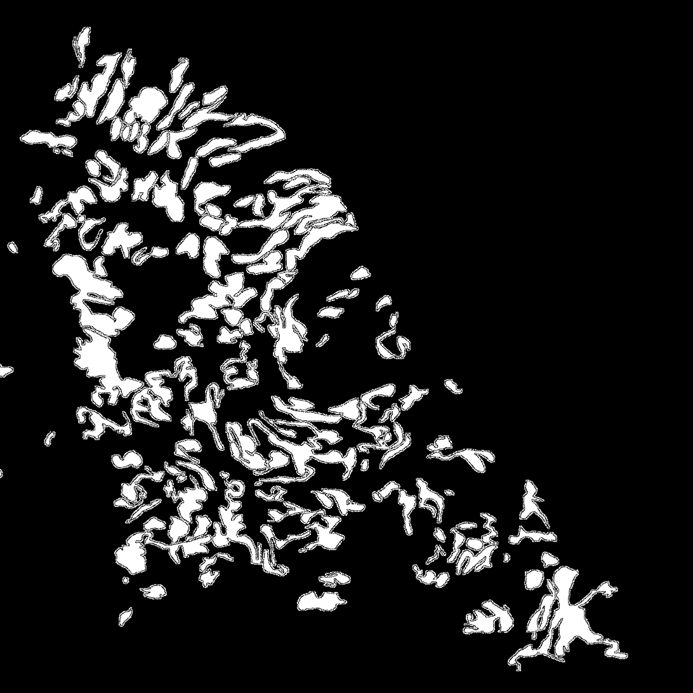}
    \end{minipage} \\
    \noalign{\vspace{2pt}}   

    \raisebox{-0.5\totalheight}{\rotatebox{90}{\scriptsize\textbf{WSI Sample 2}}} &
    \begin{minipage}{\imgw}
        \includegraphics[width=\linewidth]{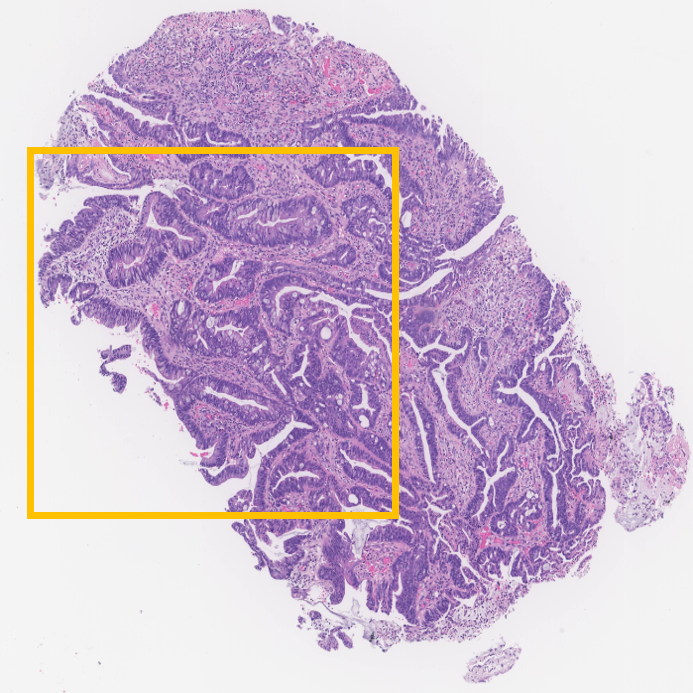}\\[\gap]
        \includegraphics[width=\linewidth]{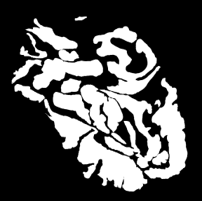}
    \end{minipage} &
    \begin{minipage}{\imgw}
        \includegraphics[width=\linewidth]{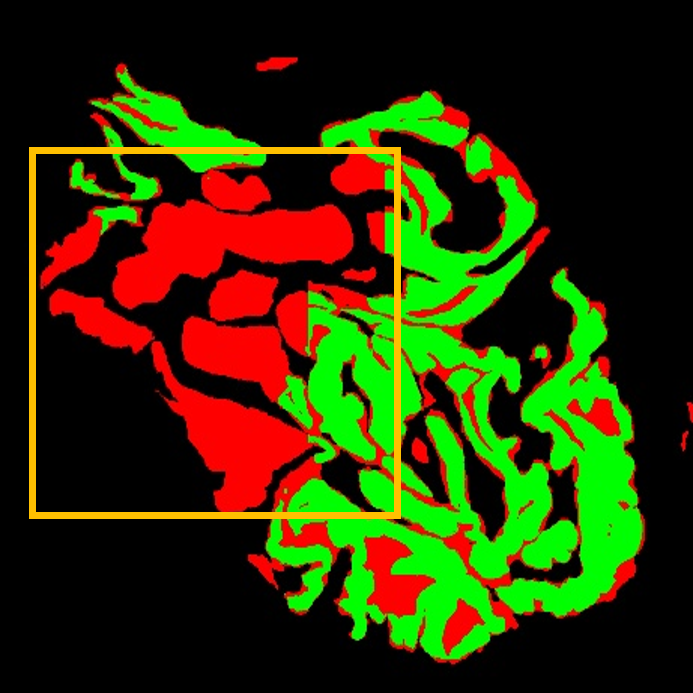}\\[\gap]
        \includegraphics[width=\linewidth]{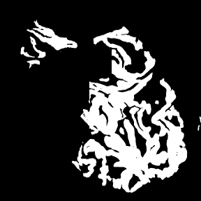}
    \end{minipage} &
    \begin{minipage}{\imgw}
        \includegraphics[width=\linewidth]{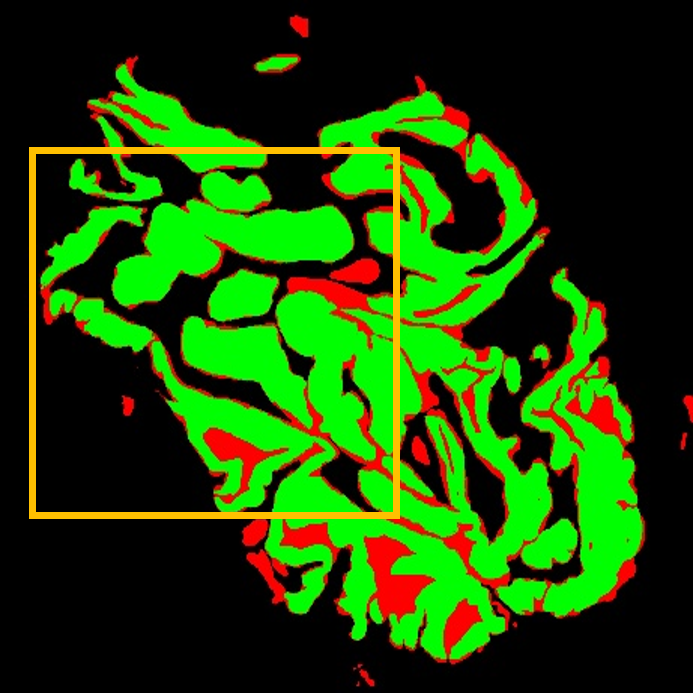}\\[\gap]
        \includegraphics[width=\linewidth]{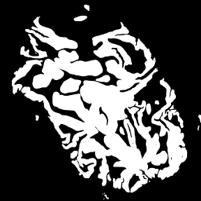}
    \end{minipage} &
    \begin{minipage}{\imgw}
        \includegraphics[width=\linewidth]{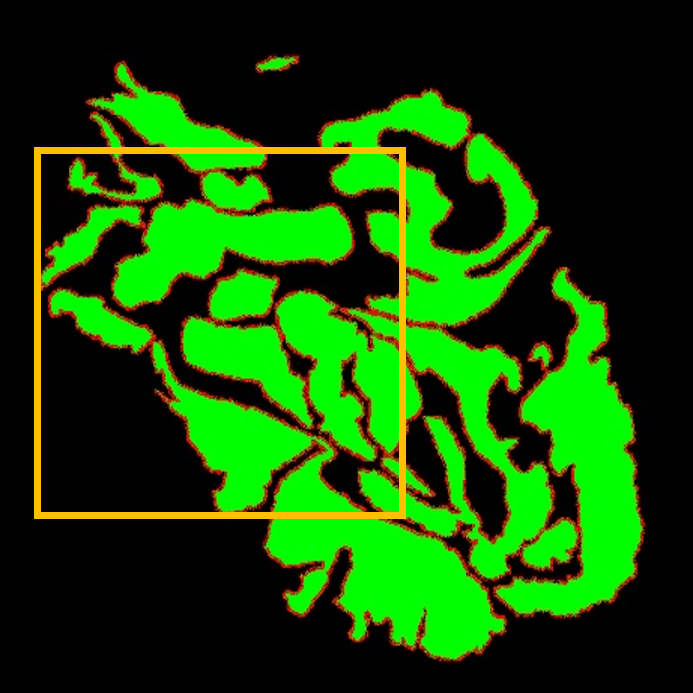}\\[\gap]
        \includegraphics[width=\linewidth]{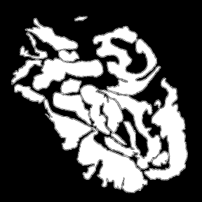}
    \end{minipage} \\

\end{tabular}%
}

\caption{\small
    \textbf{Qualitative comparison on Digestpath WSI test samples.}
    Each column displays vertically stacked pairs of (a) the input image and ground-truth annotation, followed by the error overlays and binary predictions for (b) SegFormer, (c) EViT-UNet, and (d) our proposed SAGE-ConvNeXt+ViT-UNet. WSI Sample 1 (top rows) presents branching tissue structures requiring fine local boundary delineation, while WSI Sample 2 (bottom rows) depicts a dense, uniform tissue mass prone to patch-level attention collapse in pure transformer architectures. Green areas denote correct predictions, and red areas denote errors.}
\label{fig:qualitative_results_colon}

\end{figure}

Figure~\ref{fig:qualitative_results} provides qualitative evidence supporting the improvements reported in Table~\ref{tab:baseline_results} and Table~\ref{tab:sota_results}. On GlaS Test A, SAGE-ConvNeXt+ViT-UNet produces cleaner gland boundaries and fewer false positives compared to TransUNet and EViT-UNet. The performance gap increases on the more challenging Test B sample, where competing methods exhibit pronounced stromal over-segmentation and occasional gland merging, whereas SAGE maintains gland separation and boundary fidelity. These visual trends align with the quantitative gains and indicate enhanced robustness under domain shift.

To further assess deployment behavior, Figure~\ref{fig:qualitative_results_colon} provides WSI-level qualitative comparisons on DigestPath. In WSI Sample 1, which features branching glandular structures, SAGE-ConvNeXt+ViT-UNet more accurately follows thin lesion boundaries and produces fewer false positives than SegFormer and EViT-UNet. In the denser WSI Sample 2, competing models display spillover into background tissue and partial region collapse, while SAGE preserves contiguous lesion topology with cleaner boundaries. These WSI-level findings are consistent with the quantitative improvements reported in Table~\ref{tab:digestpath_sota_results}, indicating that dynamic routing enhances stability beyond patch-level predictions.

\section{Conclusion}
\label{sec:conclusion}
We introduced SAGE (Shape-Adapting Gated Experts), a dynamic and backbone-agnostic framework for histopathology image segmentation that replaces static CNN-Transformer fusion with hierarchical expert routing and a Shape-Adapting Hub to adapt computation to input morphology; across EBHI, GlaS, and DigestPath, SAGE consistently outperforms backbone-matched baselines and recent methods at both patch and WSI levels, while offering interpretable routing behavior that clarifies expert specialization, and future work will extend evaluation to broader clinical settings and additional dense prediction tasks.

\section*{Acknowledgements}
We gratefully acknowledge The University of Texas at Austin for supporting this research, and Trivita AI and AI VIET NAM for providing the GPU computing resources essential to this work.

{
    \small
    \bibliographystyle{ieeenat_fullname}
    \bibliography{main}
}
\appendix
\clearpage
\hypersetup{pageanchor=false}
\setcounter{page}{1}
\maketitlesupplementary

\section{Semantic Affinity Routing Analysis}
Quantitative analysis of the utilization of Mixture of Experts for heterogeneous inputs reveals a fair utilization of Mixture of Experts. Analysis of the heatmap for activation of the model and attention for the expert map reveals that Semantic Affinity Routing (SAR) achieves a structured allocation of tokens to the experts, ensuring that there is no routing collapse, which is a common failure mode for sparse Mixture of Experts.

All visualizations for the routing were produced using the $GlaS$ test $A$ set~\cite{sirinukunwattana2016glandsegmentationcolonhistology} with the SAGE-ConvNeXt+ViT-UNet model and $K = 4$.

\begin{figure}[H]
    \centering
    \includegraphics[width=\linewidth]{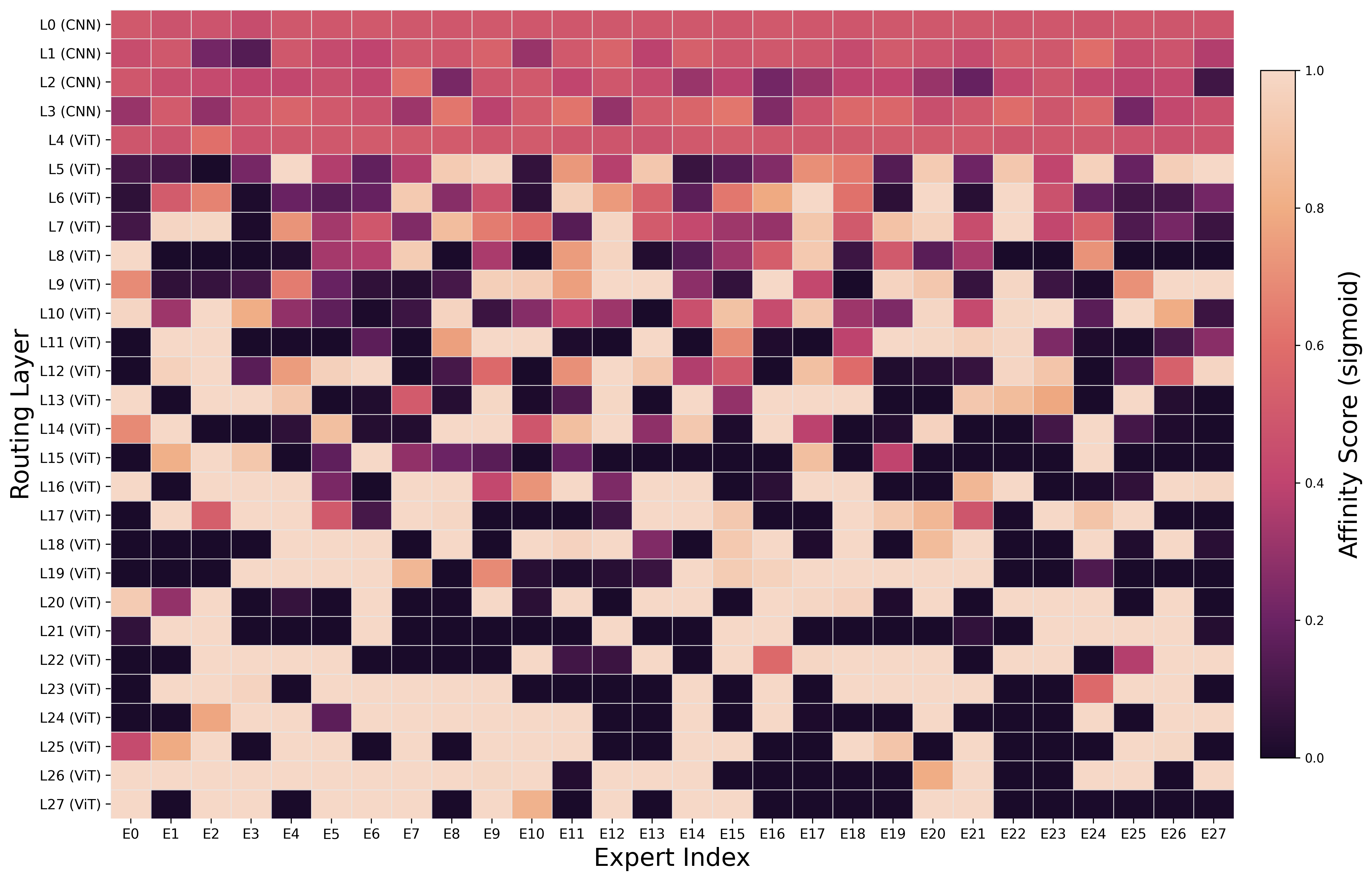}
    \caption{
        \textbf{Normalized affinity score heatmap.} The visualization shows the normalized affinity scores, which are the gating probabilities per expert per layer, with a color map ranging from red (low affinity) to dark green (high affinity). The rows in this heatmap represent the model's layer, where $L0$ to $L3$ represent the CNN layers, followed by $L4$ to $L27$, which represent the Vision Transformer layers. The columns represent the 28 experts in the expert pool, where $E0$ to $E27$ represent each individual expert.}
    \label{fig:affinity}
\end{figure}

\begin{figure} [t]
    \centering
    \includegraphics[width=\linewidth]{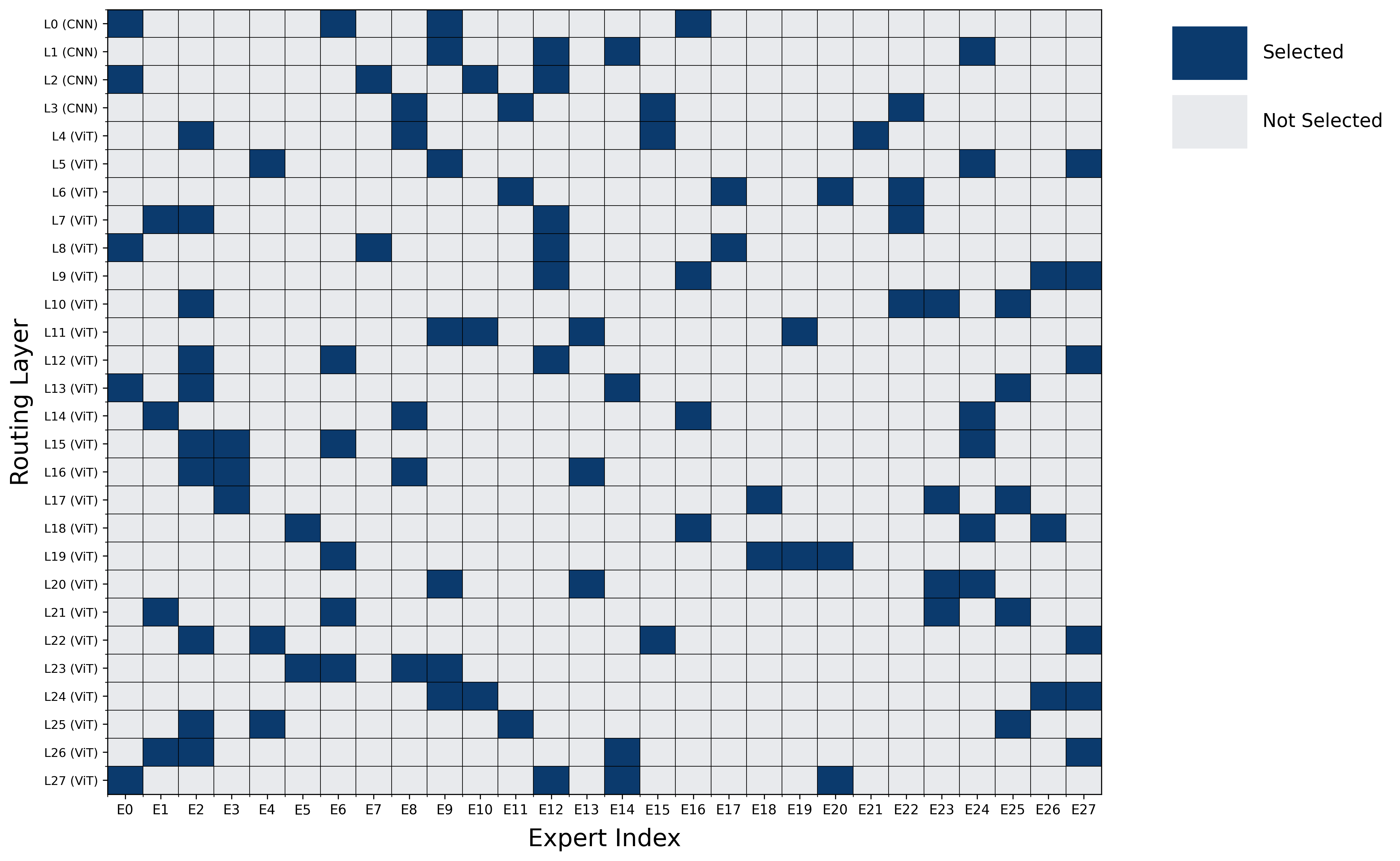}
    \caption{\textbf{Top-$K$ activation map.} The binary heatmap illustrates the routing decisions, where $K$ is equal to 4, across the 28 layers in the model (rows: $L0$ to $L3$, which represent the CNN layers; $L4$ to $L27$, which represent the Vision Transformer layers) and the 28 experts in the shared pool (columns: $E0$ to $E27$). The activation of an expert in the top K, at least for one token in the layer after processing the entire batch, is represented by blue, while empty boxes indicate no activation.}
    \label{fig:routing}
\end{figure}

\begin{figure*}[t]
    \centering
    \includegraphics[width=0.9\linewidth]{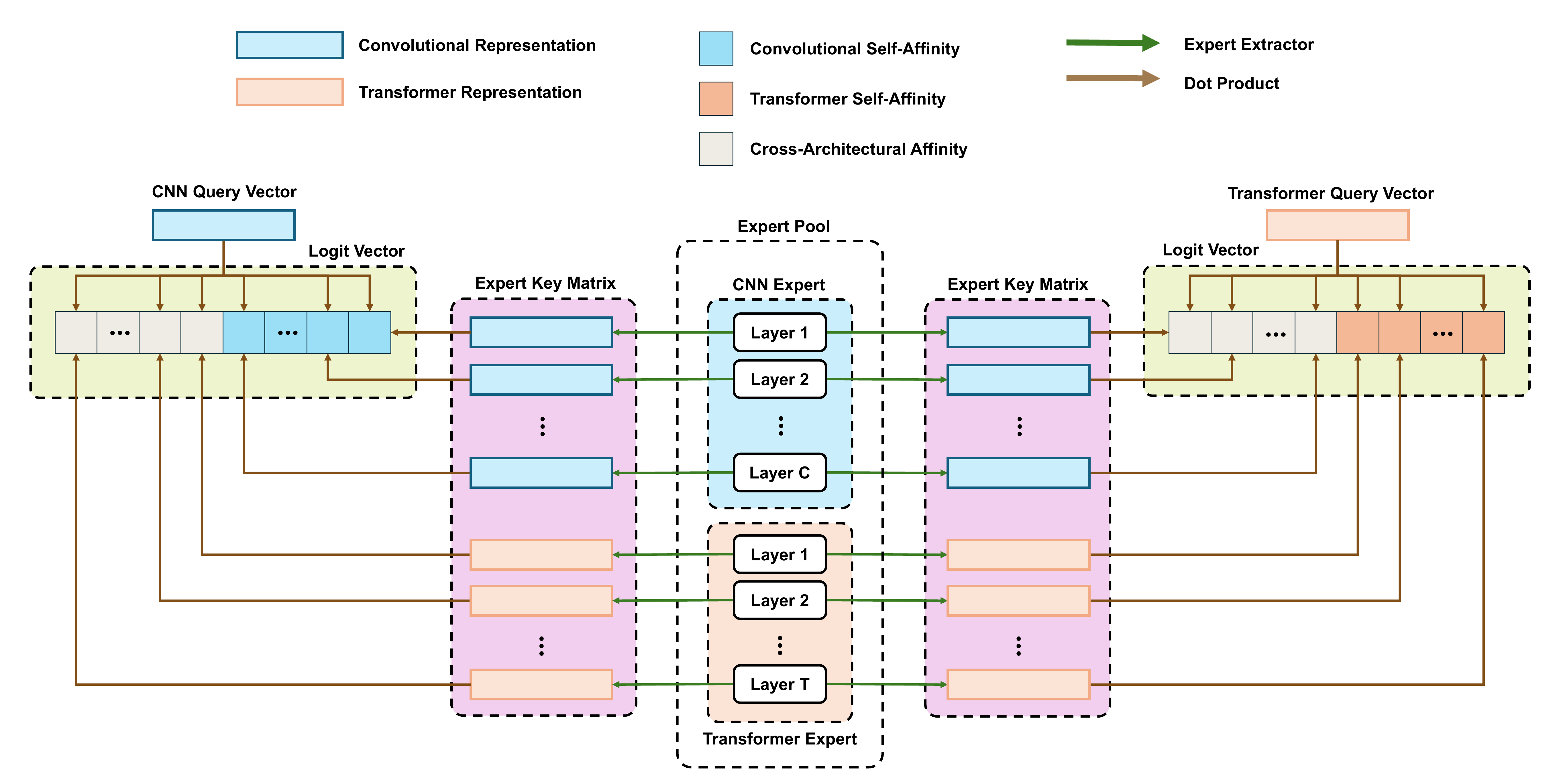}
    \caption{\textbf{Diagnostic view of SAR affinity decomposition.} Given CNN and Transformer query vectors, SAR computes per-expert routing logits by dot-product matching against expert keys extracted from both CNN and Transformer experts. The colored components separate CNN self-affinity, Transformer self-affinity, and cross-architectural affinity contributions, which are aggregated before group-level logit modulation and Top-$K$ selection.}
    \label{fig:sar_supp}
\end{figure*}

Figure~\ref{fig:affinity} shows a non-uniform affinity landscape, indicating layer-dependent expert specialization rather than routing collapse. No expert remains consistently dominant across all layers, and most experts exhibit alternating high/low affinity bands over depth. Stronger contrast appears in the early CNN-to-shallow ViT transition (approximately $L1$--$L4$), while many middle Transformer layers are closer to moderate values, with localized high-affinity reactivation in later layers. Overall, the pattern supports depth-dependent specialization rather than uniform expert usage.

In addition to the global balance observation, Figure~\ref{fig:affinity} reveals clear layer-wise variation. The early CNN and shallow ViT layers show sharper affinity contrast across experts, whereas many middle Transformer layers appear more centered around moderate affinity values. Selected late layers still present localized high-affinity experts, suggesting that specialization is redistributed across depth rather than monotonically increasing toward deeper layers.

Figure~\ref{fig:routing} complements this view by showing activation sparsity patterns per layer at $K=4$. Early CNN layers activate fewer experts per batch, whereas deeper Transformer layers trigger broader expert sets, matching their higher semantic complexity.

Taken together, the two diagnostics indicate that SAR does not behave uniformly across depth: specialization increases in later stages, while early stages remain more shared. This depth-dependent behavior is the key supplementary finding from the routing visualizations.
Figure~\ref{fig:sar_supp} provides a complementary decomposition view, showing how CNN self-affinity, Transformer self-affinity, and cross-architectural affinity jointly contribute to the final Top-$K$ routing decisions.

\section{Group-Level Gate Analysis}
\begin{figure*}[htp]
\centering
\includegraphics[width=\linewidth]{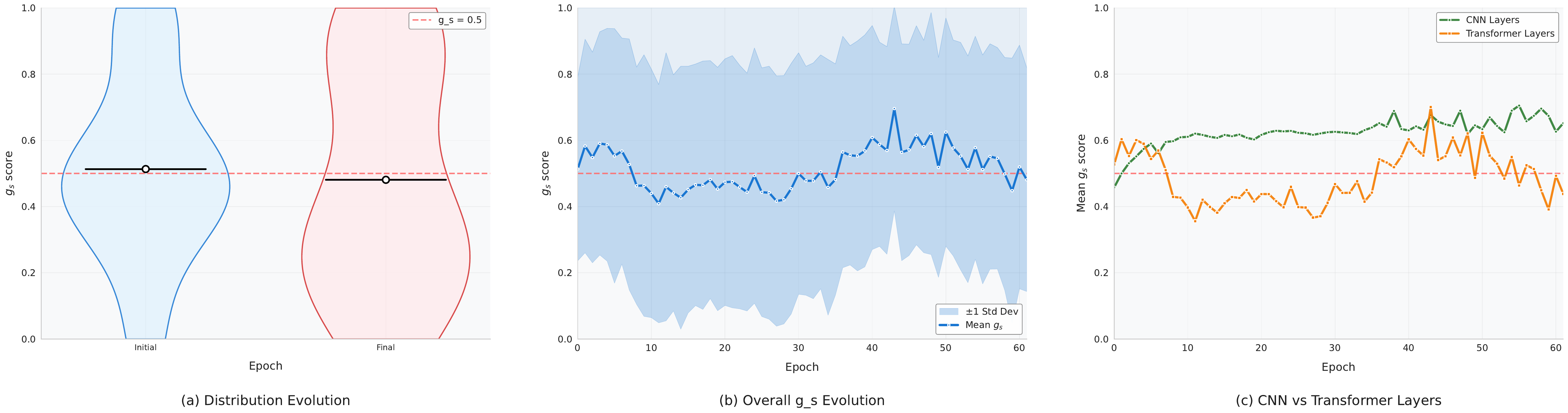}
\caption{\textbf{Group-level gate analysis during training.} The figure illustrates the evolution of $g_s$, where higher values indicate a preference for shared experts and lower values indicate a preference for fine-grained experts. \textbf{(a)} Distribution of $g_s$ at the start and end of training. \textbf{(b)} Mean $g_s$ with standard deviation across 60 epochs, with a neutral reference at $0.5$. \textbf{(c)} Mean $g_s$ by architecture type (CNN versus Transformer), showing that CNN layers remain higher while Transformer layers stay closer to the neutral region.}
\label{fig:gs_evolution}
\end{figure*}

To supplement the mechanistic evidence presented in the main paper, this section visualizes the evolution of the group-level gate $g_s$ on GlaS~\cite{sirinukunwattana2016glandsegmentationcolonhistology} during stage-2 training (Figure~\ref{fig:gs_evolution}). The hierarchical gate's formulation is identical to that in the main paper; the present analysis is restricted to its empirical behavior.

\noindent\textbf{Distribution Shift Across Training.} Figure~\ref{fig:gs_evolution}(a) demonstrates that the final distribution is broader and more polarized than the initial distribution, with substantial mass in both low-$g_s$ and high-$g_s$ regions. This observation indicates that training enhances routing selectivity at the group level, rather than converging to a narrow unimodal regime.

\noindent\textbf{Dynamic Load Balancing.} Figure~\ref{fig:gs_evolution}(b) shows that the mean $g_s$ oscillates around the neutral value ($0.5$): it decreases below $0.5$ during early and mid epochs, increases above $0.5$ in later epochs, and stabilizes near $0.5$ at the end of training. The broad standard deviation band reflects considerable sample-wise heterogeneity throughout training, consistent with input-dependent switching between shared and fine-grained experts.

\noindent\textbf{Architectural Role Specialization.} Figure~\ref{fig:gs_evolution}(c) reveals a consistent architectural gap, with CNN layers exhibiting higher $g_s$ values than Transformer layers.

\begin{itemize}
\item \textbf{CNN Layers:} Higher $g_s$ indicates stronger reliance on shared experts, which aligns with the extraction of low-level structural features.

\item \textbf{Transformer Layers:} Lower and more variable $g_s$ values suggest a more frequent preference for fine-grained experts, with periodic shifts back toward neutrality as training advances.

\end{itemize}

These results support the intended division of labor within the hybrid encoder. CNN layers are more oriented toward shared experts, whereas Transformer layers exhibit a tendency toward mixed routing.

\section{Shape-Adapting Hub Analysis}

\begin{figure*}[htp]
\centering
\includegraphics[width=0.9\linewidth]{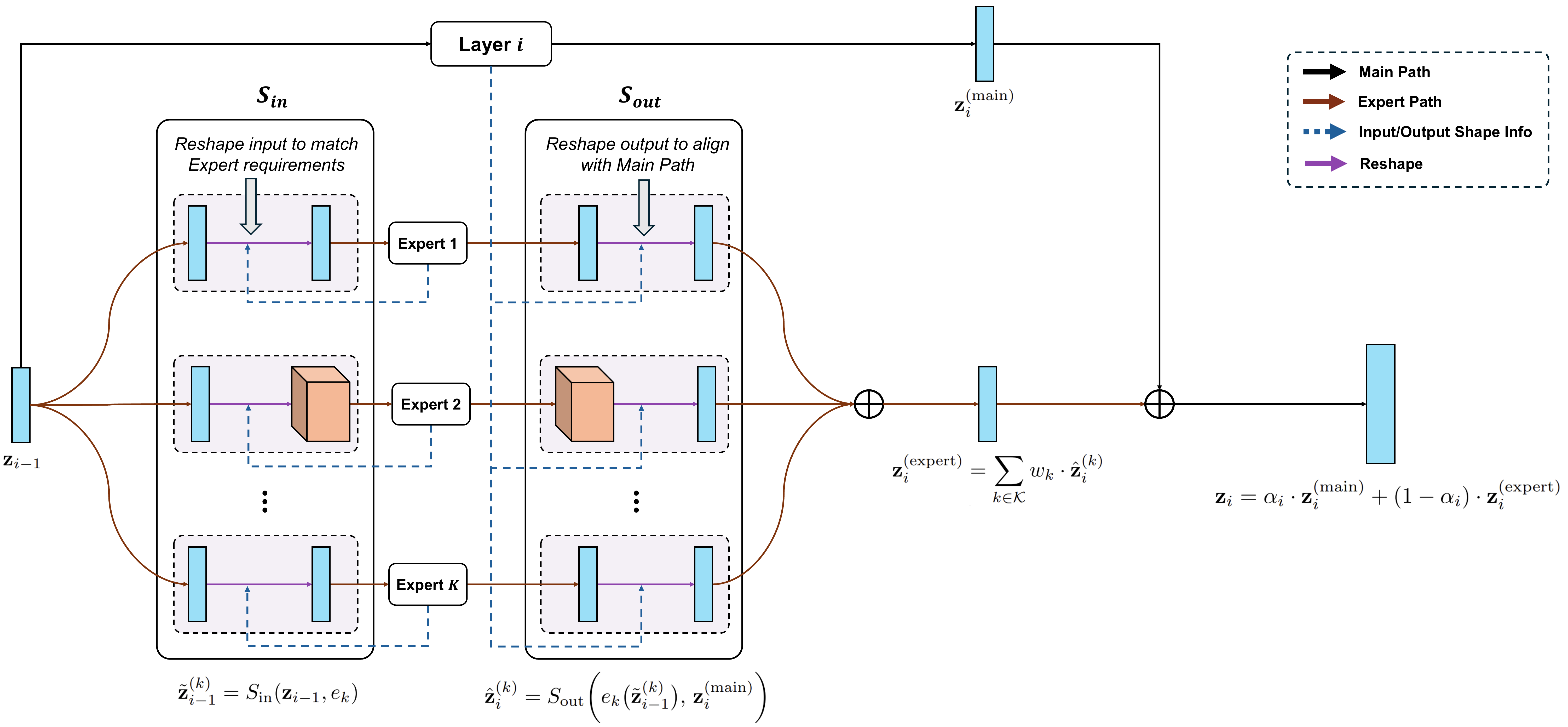}
\caption{\textbf{Detailed execution view of the Shape-Adapting Hub (SA-Hub).} The diagram highlights the adaptation path for each expert, which includes input reshaping, expert execution, and output realignment. This process is followed by weighted aggregation across activated experts and gated fusion with the main branch. Blue dashed arrows indicate the flow of shape metadata, while purple arrows represent reshape operations.}
\label{fig:sahub}
\end{figure*}

Figure~\ref{fig:sahub} provides an implementation-level perspective of the SA-Hub and highlights three operational details that are essential for interpreting routing behavior.

First, shape adaptation is performed for each activated expert. Each selected expert receives a dedicated input-side reshape and output-side realignment, enabling heterogeneous expert blocks to operate within the same routed layer.

Second, the diagram distinguishes between expert selection and expert contribution. Top-$K$ routing identifies the active experts, while subsequent weighting and fusion stages determine the extent to which each active expert influences the final representation.

Third, the dashed metadata paths indicate that adaptation is conditioned on shape or interface information rather than a single fixed reshape rule. This perspective clarifies why runtime increases with the number of activated experts, even when parameter growth remains modest.

Overall, the SA-Hub perspective illustrated in Figure~\ref{fig:sahub} elucidates the integration of heterogeneous experts into a shape-consistent representation pathway and offers a mechanistic explanation for the observed trade-off between routing flexibility and inference cost.

\section{Ablation Studies}
\label{sec:ablation}

\begin{figure*}[t] 
    \centering
    \begin{subfigure}[b]{0.48\textwidth}
        \centering
        \includegraphics[width=\linewidth]{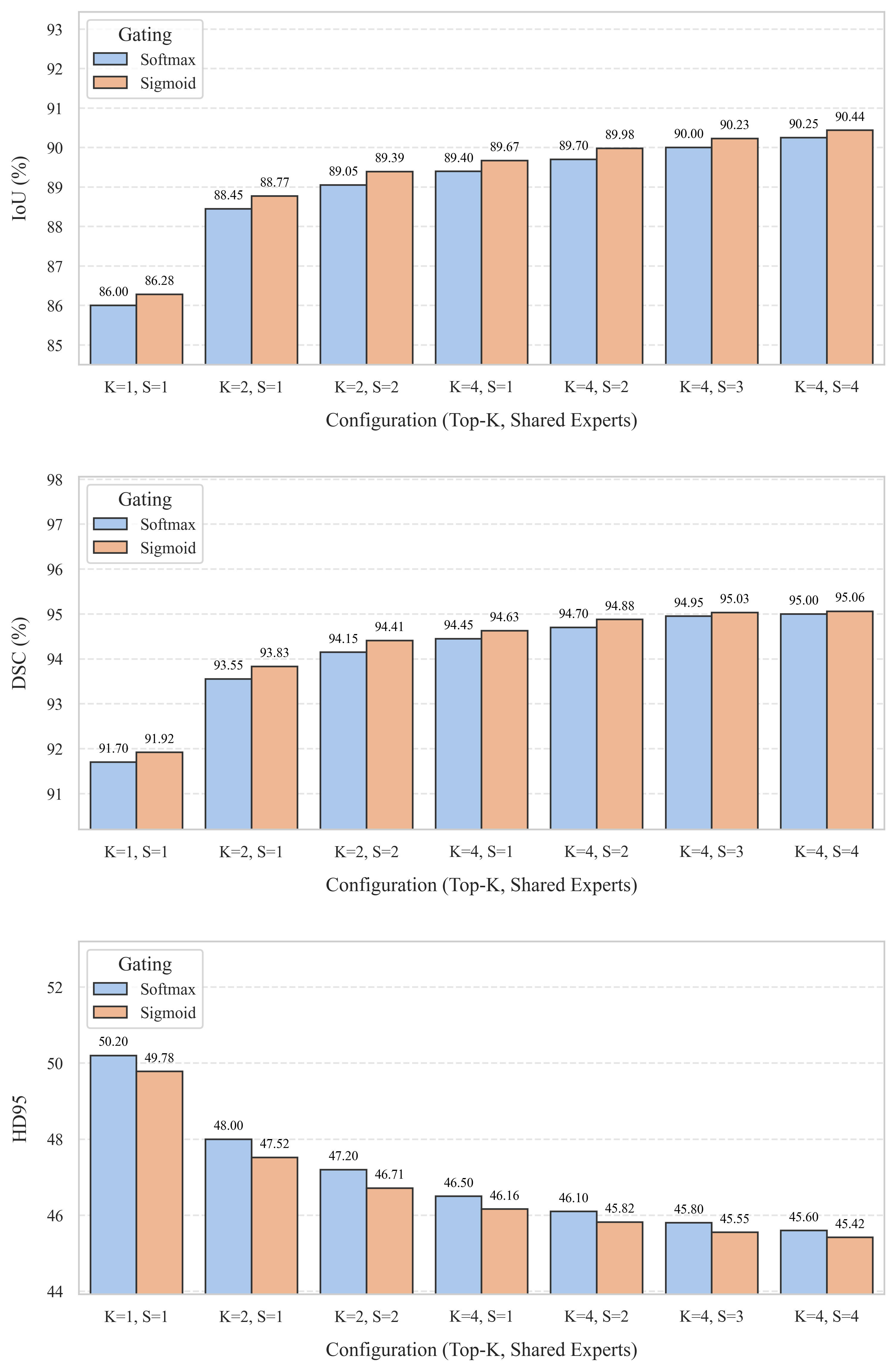}
        \caption{Without Logit Modification}
        \label{fig:ablation_off}
    \end{subfigure}
    \hfill
    \begin{subfigure}[b]{0.48\textwidth}
        \centering
        \includegraphics[width=\linewidth]{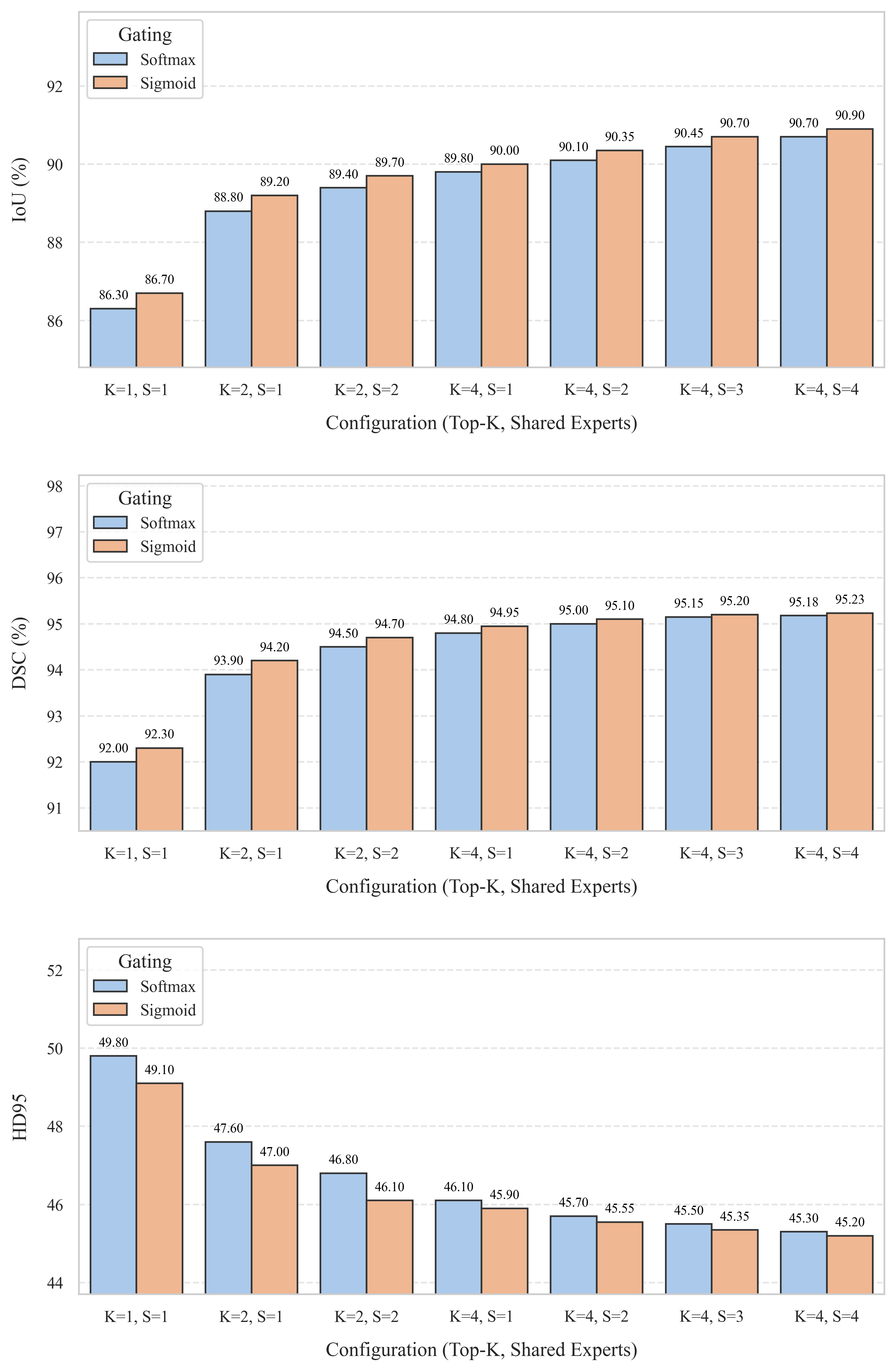}
        \caption{With Logit Modification}
        \label{fig:ablation_on}
    \end{subfigure}
    \caption{\textbf{Ablation of gating strategies and expert capacity.} Comparison of sigmoid vs softmax gating with and without logit modification across varying Top-$K$ ($K$) and shared experts ($S$).}
    \label{fig:ablation_combined}
\end{figure*}

We investigate the impact of gating strategies and expert capacity on the EBHI dataset~\cite{HU2023102534}. Figure~\ref{fig:ablation_combined} summarizes the trend, and Table~\ref{tab:ablation_scores} reports the exact scores.

\begin{table*}[t]
\centering
\caption{\textbf{Unified ablation on EBHI with and without logit modulation.} This table compares matched routing configurations under both settings. Higher values indicate better performance for Acc, IoU, DSC, and BF1, while lower values are preferable for HD95. \colorbox{BestColor}{\textcolor{BestTextColor}{\textbf{Best}}} and \colorbox{SecondBestColor}{\textcolor{SecondBestTextColor}{Second}} denote the best and second-best results within each modulation regime. The rightmost gain column reports $(\text{With}-\text{Without})$ to explicitly quantify the benefit of modulation.}
\label{tab:ablation_scores}
\setlength{\tabcolsep}{2.8pt}
\small
\resizebox{\textwidth}{!}{%
\begin{tabular}{@{}ccc|ccccc|ccccc|ccccc@{}}
\toprule
\multicolumn{3}{c|}{\textbf{Config}} & \multicolumn{5}{c|}{\textbf{Without Logit Modulation}} & \multicolumn{5}{c|}{\textbf{With Logit Modulation}} & \multicolumn{5}{c}{\textbf{Gain (With-Without)}} \\
\cmidrule(lr){1-3} \cmidrule(lr){4-8} \cmidrule(lr){9-13} \cmidrule(lr){14-18}
\textbf{Gating} & \textbf{$K$} & \textbf{$S$} & \textbf{Acc} $\uparrow$ & \textbf{IoU} $\uparrow$ & \textbf{DSC} $\uparrow$ & \textbf{HD95} $\downarrow$ & \textbf{BF1} $\uparrow$ & \textbf{Acc} $\uparrow$ & \textbf{IoU} $\uparrow$ & \textbf{DSC} $\uparrow$ & \textbf{HD95} $\downarrow$ & \textbf{BF1} $\uparrow$ & $\Delta$\textbf{Acc} $\uparrow$ & $\Delta$\textbf{IoU} $\uparrow$ & $\Delta$\textbf{DSC} $\uparrow$ & $\Delta$\textbf{HD95} $\downarrow$ & $\Delta$\textbf{BF1} $\uparrow$ \\
\midrule
Softmax & 1 & 1 & 88.85 & 86.00 & 91.70 & 50.20 & 49.70 & 89.10 & 86.30 & 92.00 & 49.80 & 50.20 & \textcolor{BestTextColor}{+0.25} & \textcolor{BestTextColor}{+0.30} & \textcolor{BestTextColor}{+0.30} & \textcolor{BestTextColor}{-0.40} & \textcolor{BestTextColor}{+0.50} \\
Sigmoid & 1 & 1 & 89.07 & 86.28 & 91.92 & 49.78 & 50.18 & 89.40 & 86.70 & 92.30 & 49.10 & 51.00 & \textcolor{BestTextColor}{+0.33} & \textcolor{BestTextColor}{+0.42} & \textcolor{BestTextColor}{+0.38} & \textcolor{BestTextColor}{-0.68} & \textcolor{BestTextColor}{+0.82} \\
\midrule
Softmax & 2 & 1 & 91.95 & 88.45 & 93.55 & 48.00 & 53.60 & 92.20 & 88.80 & 93.90 & 47.60 & 54.10 & \textcolor{BestTextColor}{+0.25} & \textcolor{BestTextColor}{+0.35} & \textcolor{BestTextColor}{+0.35} & \textcolor{BestTextColor}{-0.40} & \textcolor{BestTextColor}{+0.50} \\
Sigmoid & 2 & 1 & 92.21 & 88.77 & 93.83 & 47.52 & 54.12 & 92.55 & 89.20 & 94.20 & 47.00 & 55.10 & \textcolor{BestTextColor}{+0.34} & \textcolor{BestTextColor}{+0.43} & \textcolor{BestTextColor}{+0.37} & \textcolor{BestTextColor}{-0.52} & \textcolor{BestTextColor}{+0.98} \\
\midrule
Softmax & 2 & 2 & 92.60 & 89.05 & 94.15 & 47.20 & 55.40 & 92.90 & 89.40 & 94.50 & 46.80 & 56.00 & \textcolor{BestTextColor}{+0.30} & \textcolor{BestTextColor}{+0.35} & \textcolor{BestTextColor}{+0.35} & \textcolor{BestTextColor}{-0.40} & \textcolor{BestTextColor}{+0.60} \\
Sigmoid & 2 & 2 & 92.88 & 89.39 & 94.41 & 46.71 & 55.92 & 93.25 & 89.70 & 94.70 & 46.10 & 56.90 & \textcolor{BestTextColor}{+0.37} & \textcolor{BestTextColor}{+0.31} & \textcolor{BestTextColor}{+0.29} & \textcolor{BestTextColor}{-0.61} & \textcolor{BestTextColor}{+0.98} \\
\midrule
Softmax & 4 & 1 & 93.10 & 89.40 & 94.45 & 46.50 & 56.60 & 93.45 & 89.80 & 94.80 & 46.10 & 57.10 & \textcolor{BestTextColor}{+0.35} & \textcolor{BestTextColor}{+0.40} & \textcolor{BestTextColor}{+0.35} & \textcolor{BestTextColor}{-0.40} & \textcolor{BestTextColor}{+0.50} \\
Sigmoid & 4 & 1 & 93.34 & 89.67 & 94.63 & 46.16 & 56.97 & 93.70 & 90.00 & 94.95 & 45.90 & 57.60 & \textcolor{BestTextColor}{+0.36} & \textcolor{BestTextColor}{+0.33} & \textcolor{BestTextColor}{+0.32} & \textcolor{BestTextColor}{-0.26} & \textcolor{BestTextColor}{+0.63} \\
\midrule
Softmax & 4 & 2 & 93.30 & 89.70 & 94.70 & 46.10 & 57.20 & 93.62 & 90.10 & 95.00 & 45.70 & 57.80 & \textcolor{BestTextColor}{+0.32} & \textcolor{BestTextColor}{+0.40} & \textcolor{BestTextColor}{+0.30} & \textcolor{BestTextColor}{-0.40} & \textcolor{BestTextColor}{+0.60} \\
Sigmoid & 4 & 2 & 93.58 & 89.98 & 94.88 & 45.82 & 57.48 & 93.85 & 90.35 & 95.10 & 45.55 & 58.00 & \textcolor{BestTextColor}{+0.27} & \textcolor{BestTextColor}{+0.37} & \textcolor{BestTextColor}{+0.22} & \textcolor{BestTextColor}{-0.27} & \textcolor{BestTextColor}{+0.52} \\
\midrule
Softmax & 4 & 3 & 93.45 & 90.00 & 94.95 & 45.80 & 57.40 & 93.78 & 90.45 & 95.15 & 45.50 & 58.00 & \textcolor{BestTextColor}{+0.33} & \textcolor{BestTextColor}{+0.45} & \textcolor{BestTextColor}{+0.20} & \textcolor{BestTextColor}{-0.30} & \textcolor{BestTextColor}{+0.60} \\
Sigmoid & 4 & 3 & \cellcolor{SecondBestColor}\textcolor{SecondBestTextColor}{93.66} & 90.23 & \cellcolor{SecondBestColor}\textcolor{SecondBestTextColor}{95.03} & \cellcolor{SecondBestColor}\textcolor{SecondBestTextColor}{45.55} & \cellcolor{SecondBestColor}\textcolor{SecondBestTextColor}{57.56} & \cellcolor{SecondBestColor}\textcolor{SecondBestTextColor}{93.96} & \cellcolor{SecondBestColor}\textcolor{SecondBestTextColor}{90.70} & \cellcolor{SecondBestColor}\textcolor{SecondBestTextColor}{95.20} & \cellcolor{SecondBestColor}\textcolor{SecondBestTextColor}{45.35} & \cellcolor{SecondBestColor}\textcolor{SecondBestTextColor}{58.05} & \textcolor{BestTextColor}{+0.30} & \textcolor{BestTextColor}{+0.47} & \textcolor{BestTextColor}{+0.17} & \textcolor{BestTextColor}{-0.20} & \textcolor{BestTextColor}{+0.49} \\
\midrule
Softmax & 4 & 4 & 93.55 & \cellcolor{SecondBestColor}\textcolor{SecondBestTextColor}{90.25} & 95.00 & 45.60 & 57.50 & 93.88 & \cellcolor{SecondBestColor}\textcolor{SecondBestTextColor}{90.70} & 95.18 & 45.30 & \cellcolor{SecondBestColor}\textcolor{SecondBestTextColor}{58.05} & \textcolor{BestTextColor}{+0.33} & \textcolor{BestTextColor}{+0.45} & \textcolor{BestTextColor}{+0.18} & \textcolor{BestTextColor}{-0.30} & \textcolor{BestTextColor}{+0.55} \\
Sigmoid & 4 & 4 & \cellcolor{BestColor}\textcolor{BestTextColor}{93.74} & \cellcolor{BestColor}\textcolor{BestTextColor}{90.44} & \cellcolor{BestColor}\textcolor{BestTextColor}{95.06} & \cellcolor{BestColor}\textcolor{BestTextColor}{45.42} & \cellcolor{BestColor}\textcolor{BestTextColor}{57.63} & \cellcolor{BestColor}\textcolor{BestTextColor}{94.03} & \cellcolor{BestColor}\textcolor{BestTextColor}{90.90} & \cellcolor{BestColor}\textcolor{BestTextColor}{95.23} & \cellcolor{BestColor}\textcolor{BestTextColor}{45.20} & \cellcolor{BestColor}\textcolor{BestTextColor}{58.10} & \textcolor{BestTextColor}{+0.29} & \textcolor{BestTextColor}{+0.46} & \textcolor{BestTextColor}{+0.17} & \textcolor{BestTextColor}{-0.22} & \textcolor{BestTextColor}{+0.47} \\
\bottomrule
\end{tabular}%
}
\normalsize
\setlength{\tabcolsep}{6pt}
\end{table*}

\noindent\textbf{Unified Comparison (Without vs. With Logit Modulation).} Table~\ref{tab:ablation_scores} is structured to facilitate row-wise comparisons under identical $(\text{gating}, K, S)$ settings. In all seven configuration pairs, logit modulation consistently increases Acc, IoU, DSC, and BF1, while reducing HD95 for both sigmoid and softmax gating.

\noindent\textbf{Gating Mechanism.} The selection of the gating function is critical for the router's capacity to address complex tissue morphologies. As shown in Table~\ref{tab:ablation_scores}, sigmoid gating consistently outperforms softmax gating in both regimes (with and without logit modulation) across all evaluated settings. In the most expansive configuration ($K=4$, $S=4$), the sigmoid router achieves a DSC of $95.06\%$ without logit modulation and $95.23\%$ with modulation, while the softmax router attains $95.00\%$ and $95.18\%$, respectively. This trend aligns with previous findings in sigmoid-based routing and attention mechanisms~\cite{sigmoidGatingMoE, sigmoid_attention}. The non-competitive nature of sigmoid gating is particularly beneficial for histopathology image segmentation, where the integration of multiple distinct feature extractors is often required due to the presence of diverse and overlapping tissues.

\noindent\textbf{Effect of Logit Modulation.} The performance improvements from logit modulation are systematic and increase additively with greater routing capacity. For example, in the optimal sigmoid configuration ($K=4$, $S=4$), enabling logit modulation raises the DSC from $95.06\%$ to $95.23\%$ ($+0.17\%$) and the boundary-aware BF1 score from $57.63$ to $58.10$ ($+0.47$). It also reduces the HD95 distance from $45.42$ to $45.20$ ($-0.22$), indicating enhanced structural boundary precision.

\noindent\textbf{Expert Capacity ($K$ and $S$).} In both modulation regimes, increasing the number of selected experts ($K$) yields the largest performance improvements. After achieving a high level of specialization ($K=4$), further increasing the shared expert pool ($S$) results in smaller but consistent incremental gains. These findings indicate that $K$ primarily enhances the model's ability to process diverse inputs, while $S$ serves as a stable common-knowledge anchor that mitigates feature fragmentation at higher routing capacities.

Based on these comprehensive results, the $K=4$ and $S=4$ architecture with sigmoid gating and logit modulation is adopted as the reference configuration, as it offers the optimal balance between specialized routing and stable feature aggregation.

\end{document}